\PassOptionsToPackage{pdfpagelabels=false}{hyperref} 
\documentclass[fleqn,usenatbib]{mnras}

\usepackage{newtxtext,newtxmath}

\usepackage[T1]{fontenc}
\usepackage{ae,aecompl}

\usepackage{graphicx}	
\usepackage{amsmath}	

\usepackage{mathtools}
\usepackage{natbib}
\usepackage{color}
\usepackage{hyperref}
\usepackage{comment}
\usepackage{multirow}
\usepackage{tabularx}

\usepackage{ulem}

\usepackage{lineno}

\definecolor{Green}{rgb}{0,0,0.9}
\definecolor{ao}{rgb}{0.0, 0.5, 0.0}

\newcommand{\grizy}{$g, r, i, z, Y$ }

\newcommand{\RM}{redMaPPer}

\defcitealias{Zhang2019}{Z19}
\defcitealias{Zibetti2005}{Z05}

\usepackage{eso-pic}

\AddToShipoutPictureBG*{%
  \AtPageUpperLeft{%
    \hspace{0.75\paperwidth}%
    \raisebox{-3.5\baselineskip}{%
      \makebox[0pt][l]{\textnormal{DES 2020-0527}}
}}}%

\AddToShipoutPictureBG*{%
  \AtPageUpperLeft{%
    \hspace{0.75\paperwidth}%
    \raisebox{-4.5\baselineskip}{%
      \makebox[0pt][l]{\textnormal{FERMILAB-PUB-20-205-AE-E}}
}}}%

\title[Diffuse light tracer of galaxy cluster mass]{Is diffuse intracluster light a good tracer of the galaxy cluster matter distribution?}

\author[DES Collaboration]{
\parbox{\textwidth}{
\Large
H.~Sampaio-Santos$^{1,2}$,
Y.~Zhang$^{3}$,
R.~L.~C.~Ogando$^{1,2}$,
T.~Shin$^{4}$,
Jesse~B.~Golden-Marx$^{5,6}$,
B.~Yanny$^{3}$,
K.~Herner$^{3}$,
M.~Hilton$^{7}$,
A.~Choi$^{8}$,
M.~Gatti$^{9}$,
D.~Gruen$^{10,11,12}$,
B.~Hoyle$^{13,14}$,
M.~M.~Rau$^{15}$,
J.~De~Vicente$^{16}$,
J.~Zuntz$^{17}$,
T.~M.~C.~Abbott$^{18}$,
M.~Aguena$^{19,2}$,
S.~Allam$^{3}$,
J.~Annis$^{3}$,
S.~Avila$^{20}$,
E.~Bertin$^{21,22}$,
D.~Brooks$^{23}$,
D.~L.~Burke$^{11,12}$,
M.~Carrasco~Kind$^{24,25}$,
J.~Carretero$^{9}$,
C.~Chang$^{26,27}$,
M.~Costanzi$^{28,29}$,
L.~N.~da Costa$^{2,1}$,
H.~T.~Diehl$^{3}$,
P.~Doel$^{23}$,
S.~Everett$^{30}$,
A.~E.~Evrard$^{6,31}$,
B.~Flaugher$^{3}$,
P.~Fosalba$^{32,33}$,
J.~Frieman$^{3,27}$,
J.~Garc\'ia-Bellido$^{20}$,
E.~Gaztanaga$^{32,33}$,
D.~W.~Gerdes$^{6,31}$,
R.~A.~Gruendl$^{24,25}$,
J.~Gschwend$^{2,1}$,
G.~Gutierrez$^{3}$,
S.~R.~Hinton$^{34}$,
D.~L.~Hollowood$^{30}$,
K.~Honscheid$^{8,35}$,
D.~J.~James$^{36}$,
M.~Jarvis$^{4}$,
T.~Jeltema$^{30}$,
K.~Kuehn$^{37,38}$,
N.~Kuropatkin$^{3}$,
O.~Lahav$^{23}$,
M.~A.~G.~Maia$^{2,1}$,
M.~March$^{4}$,
J.~L.~Marshall$^{39}$,
R.~Miquel$^{40,9}$,
A.~Palmese$^{3,27}$,
F.~Paz-Chinch\'{o}n$^{41,25}$,
A.~A.~Plazas$^{42}$,
E.~Sanchez$^{16}$,
B.~Santiago$^{43,2}$,
V.~Scarpine$^{3}$,
M.~Schubnell$^{31}$,
M.~Smith$^{44}$,
E.~Suchyta$^{45}$,
G.~Tarle$^{31}$,
D.~L.~Tucker$^{3}$,
T.~N.~Varga$^{13,14}$,
R.~H.~Wechsler$^{10,11,12}$
\begin{center} (DES Collaboration) \end{center}
}
\vspace{0.4cm}
\\
}

\date{Accepted XXX. Received YYY; in original form ZZZ}

\pubyear{2020}

\begin{document}
\label{firstpage}
\pagerange{\pageref{firstpage}--\pageref{lastpage}}
\maketitle

\begin{abstract}

We explore the relation between diffuse intracluster light (central galaxy included) and the galaxy cluster (baryonic and dark) matter distribution using a sample of 528 clusters at $0.2\leq z \leq 0.35$ found in the Dark Energy Survey (DES) Year 1 data. The surface brightness of the diffuse light shows an increasing dependence on cluster total mass at larger radius, and appears to be self-similar with a universal radial dependence after scaling by cluster radius. We also compare the diffuse light radial profiles to the cluster (baryonic and dark) matter distribution measured through weak lensing and find them to be comparable. The IllustrisTNG galaxy formation simulation, TNG300, offers further insight into the connection between diffuse stellar mass and cluster matter distributions -- the simulation radial profile of the diffuse stellar component does not have a similar slope with the total cluster matter content, although that of the cluster satellite galaxies does. Regardless of the radial trends, the amount of diffuse stellar mass has a low-scatter scaling relation with cluster's total mass in the simulation, out-performing the total stellar mass of cluster satellite galaxies. We conclude that there is no consistent evidence yet on whether or not diffuse light is a faithful radial tracer of the cluster matter distribution. Nevertheless, both observational and simulation results reveal that diffuse light is an excellent indicator of the cluster's total mass.

\end{abstract}




\section{Introduction}

Galaxy clusters are permeated by a diffuse component known as the intracluster light (ICL), composed of stars that do not appear to be bound to any of the galaxies in a cluster. 
The existence of ICL was first reported by \cite{Zwicky1951} almost 70 years ago, but, limited by its very low surface brightness ( measured as $\sim$ 30 mag/arcsec$^{2}$ in $r$-band, \citealt{Zhang2019}), only in the 1990's ICL started to receive wide attention due to technological advancements such as the CCD camera \citep{Uson1991,Bernstein1995,Gonzalez2000,Feldmeier2003}.

Given its low surface brightness level, diffuse intracluster light is difficult to observe, nevertheless, it is an important component of galaxy clusters. Observations, semi-analytical modeling, and simulation studies report that the ICL and cluster central galaxies may make up 10 - 50\% of the total cluster stellar light \citep[e.g.][]{Feldmeier2004,Zibetti2005,Gonzalez2007,Behroozi2013,Pillepich2014, Zhang2019}.
An interesting new perspective on intracluster light is its  connection to the cluster dark matter distribution. \cite{Montes2019} observed a striking similarity between the shape of cluster dark matter distribution and diffuse intracluster light, even more than between dark matter and intracluster gas. A possible explanation is that both dark matter and diffuse intracluster light contain collisionless particles, while the intracluster gas has  self-interaction and dissipation. Thus, diffuse intracluster light is potentially a better tracer of the cluster dark matter distribution and an alternative mass proxy for wide and deep surveys such as LSST \citep{lsst2019}.

Another evidence of the connection between dark matter and ICL was shown by \citet{Montes2018}, there, the 3-D slope of diffuse light measured in 6 clusters in the Hubble Frontier Fields follows the expected 3-D slope of a dark matter halo in IllustrisTNG. Other works also find a correlation between cluster mass (including dark matter) and the total diffuse light luminosity or stellar mass, especially at large radius \citep[e.g.,][]{Zibetti2005, 2020ApJS..247...43K,2020MNRAS.491.3751D, 2018PASJ...70S...6H, 2018MNRAS.480..521H}. 
Furthermore, \citet[][hereafter Z19]{Zhang2019} discovered that the ratio between diffuse light surface brightness and a weak-lensing measurement-based cluster mass-density model appears to be flat at cluster radius greater than 100 kpc, and that diffuse intracluster light radial profiles are self-similar. 

Recently, several groups started to analyze this elusive component in N-body simulations. \citet{Asensio2020} investigated 30 simulated galaxy clusters within a narrow range of mass ($10^{14} < M_{200}/M_{\odot} < 10^{15.4}$) in the Cluster-EAGLE suite and found that their stellar mass and total matter have similar distributions, even more than in \citet{Montes2019}. Probing a larger range of halo masses in the Horizon-AGN simulation, \citet{canas19} found that the diffuse light stellar mass fraction increases with halo mass, while its scatter decreases with mass. 

In this paper, we explore the connection between diffuse intracluster light and cluster dark matter distribution using data from the Dark Energy Survey (DES, \citealt{DES2016}), a wide-field optical imaging survey in \grizy using the 4-meter Blanco telescope and the  Dark Energy Camera \citep[DECam,][]{Flaugher15}. 
The analysis of diffuse light in galaxy clusters greatly benefits from extremely wide-field surveys like 
SDSS \citep[e.g.][hereafter Z05]{Zibetti2005} and 
DES \citepalias[e.g.][]{Zhang2019} because of their statistical power. \citetalias{Zhang2019} successfully detected the diffuse intracluster light using DES data out to a cluster radius range of 1 - 2 Mpc at redshift  $\sim$ 0.25 by averaging $\sim$ 300 clusters. 
We use the \citetalias{Zhang2019} methods and update their analysis with a larger sample (528 galaxy clusters) to examine the relation between diffuse light and galaxy cluster mass. Given the difficulties in separating intracluster light and the cluster central galaxy, we follow the convention in \cite{Pillepich2018} to analyze intracluster light and cluster central galaxy together as ``{\bf diffuse light}", while ``{\bf intracluster light}" or ICL is reserved to qualitatively describe the unbound light beyond a few tens of kiloparsecs around the galaxy cluster center. Tab.~\ref{tbl:definition} summarizes the definitions used in this paper.

This paper is organized as follows: in Sec. \ref{sec:data} we describe the DES data (e.g., images, source catalogs, and galaxy cluster catalogs) and our analysis methods. In Sec. \ref{sec:massdep} we explore how diffuse light profiles behave as a function of galaxy cluster mass. We also investigate if the profiles are self-similar, and its ratio to cluster total light. In Sec. \ref{sec:goodtracer} and \ref{sec:Simulationresult}, we explore the main question of this paper -- whether or not diffuse light can be used as a tracer of the cluster matter distribution, first, by comparing the diffuse light radial distribution to that of cluster total matter measured with weak lensing in Sec. \ref{sec:goodtracer}. Then, we analyze the diffuse light properties in the IllustrisTNG hydrodynamic simulations \citep{Pillepich2018} in Sec.~\ref{sec:Simulationresult} and compare to our measurements. Finally, we discuss and summarize the results in Sec. \ref{sec:conclusions}. In agreement with \citetalias{Zhang2019}, cosmological distances are calculated with a flat $\Lambda$CDM model with $h = 0.7$ and $\Omega_\mathrm{m}$ = 0.3.

\begin{table}
	\centering
	\caption{Nomenclature used in this paper.}
	\begin{tabularx}{0.5\textwidth}{|X|p{4cm}|}
	\hline
	\multicolumn{1}{|c|}{Name}
	& \multicolumn{1}{c|}{Definition}
	\\ \hline
    Cluster Central Galaxy, Central Galaxy, CG & The cluster central galaxy identified by the redMaPPer algorithm. Qualitatively, these names refer to the light/stellar mass contained within the inner $\sim$ 30 kpc of the galaxy center.
    \\ \hline
	Cluster Satellite Galaxies & The light or stellar mass contained in the non-central cluster galaxies, each defined within a Kron aperture observationally, or within twice the stellar half mass radius in the simulation.
	\\ \hline
    Intracluster Light, Diffuse Intracluster Light, ICL & The diffuse light beyond the outskirts of the central cluster galaxy, but not associated with any cluster satellite galaxy. Qualitatively, these names refer to the light/stellar mass not already contained in the cluster central galaxy or the cluster satellite galaxies.
	\\ \hline
	Diffuse Light, 
	Diffuse Stellar Mass & The light or stellar mass combination of intracluster light and the cluster central galaxy.
	\\ \hline
	Cluster Total Light, Total Cluster Light, Cluster Total Stellar Mass, Total Cluster Stellar Mass & Total light or stellar mass contained in the galaxy cluster within a cluster radial range specified in the context. This is the combination of diffuse light and cluster satellite galaxies.
	\\ \hline
	\end{tabularx}
	\label{tbl:definition}
\end{table}

\section{Data and methods}
\label{sec:data}

Our analysis is based on the observations collected and processed by the Dark Energy Survey\footnote{\url{https://www.darkenergysurvey.org}}. 
In this paper, we closely follow  \citetalias{Zhang2019} in terms of the adopted data products and diffuse light measurement methods. This section provides a brief review, and notes any differences from \citetalias{Zhang2019}. 

\subsection{The \RM{} cluster sample} 
\label{sec:redmapperimgcat}

\begin{figure}
	\includegraphics[width=\columnwidth]{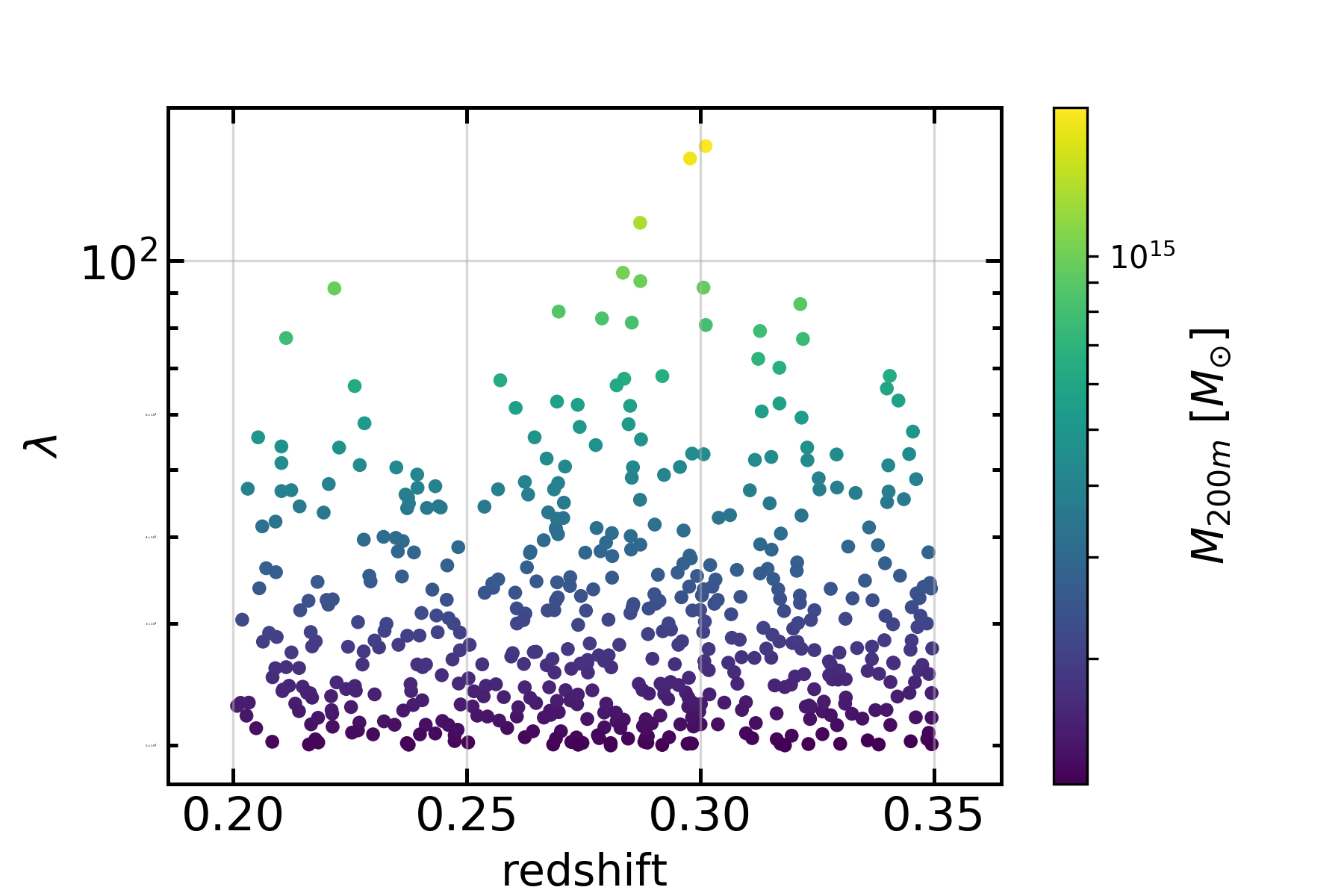}
    \caption{Richness as a function of redshift for 528 galaxy clusters. The colour represents the mean mass computed for the redMaPPer clusters. We use the mass-richness relation and the best-fit parameters reported in \citet{McClintock2019} to obtain the cluster  masses from their richnesses.}
    \label{fig:rredsh}
\end{figure}

As in \citetalias{Zhang2019}, we use the galaxy cluster sample identified by the \textbf{red}-sequence \textbf{Ma}tched-filter \textbf{P}robabilistic
\textbf{Per}colation (\RM ) algorithm (\citealt{Rykoff2014}) in DES Year 1 data. 
Each identified cluster is assigned a richness value, denoted as $\lambda$, which has been shown to be an excellent low scatter indicator of cluster mass \citep[e.g.][]{Rozo2014, Farahi2019}.
To minimize the need for applying redshift-related corrections, we only make use of the clusters in a narrow redshift range (e.g., 0.2 $\leq$ z $\leq$ 0.35). The upper redshift limit is  higher than \citetalias{Zhang2019} to match the weak lensing studies performed on the same cluster sample in \citet{McClintock2019}. We further split our sample into four richness bins: 20 $\leq$ $\lambda$ < 30, 30 $\leq$ $\lambda$ < 45, 45 $\leq$ $\lambda$ < 60 and 60 $\leq$ $\lambda$ < 150, again following the choice in \citet{McClintock2019}. Our selection ends with 538 clusters in total, 305, 149, 52, and 32 clusters in each of the respective richness bins. We use the mass-$\lambda$ relation from \citet{McClintock2019} to estimate cluster mass $M_\mathrm{200m}$, defined as the mass inside a spherical radius within which the cluster has a 200 times overdensity with respect to the universe mean matter density at the cluster's redshift. The lowest richness value from our cluster sample corresponds to a $M_\mathrm{200m}$ value of 1.2 $\times$ 10$^{14}$ M$_\odot$, while the highest richness value corresponds to a $M_\mathrm{200m}$ value of 1.8 $\times$ 10$^{15}$ M$_\odot$. We further follow up the cluster images and note 10 bad images in our cluster sample (for instance, with unmasked objects and very bright regions caused by nearby stars). We remove them from our analysis, reducing the cluster sample size to 528 in total, 297, 148, 52 and 31 clusters at 20 $\leq$ $\lambda$ < 30, 30 $\leq$ $\lambda$ < 45, 45 $\leq$ $\lambda$ < 60 and 60 $\leq$ $\lambda$ < 150, respectively. Figure \ref{fig:rredsh} shows the redshift, richness and mass distribution of those clusters.

\subsection{Light profile measurement}
\label{sec:lprof}

\begin{figure*}
	\includegraphics[width=2\columnwidth]{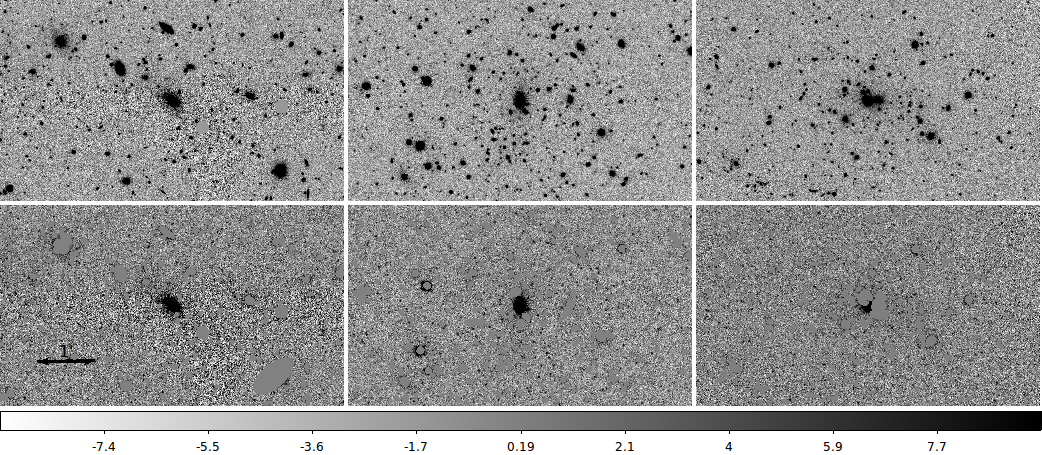}
    \caption{These figures show the $r-$band images of three galaxy clusters in our sample (top panels), and demonstrate the masking performed on each cluster (bottom panels), excluding the cluster central galaxies. Images are displayed with DS9 linear zscale. On the bottom left panel, we show a 1 arcmin scale for reference.}
    \label{fig:masking}
\end{figure*}

We make diffuse light measurements around the redMaPPer-selected central galaxies. The diffuse light in this analysis is derived from single-epoch images from the DES Year 3 processing campaign by the DES Data Management (DES DM) team \citep{Abbot18DesDr1}. 
For a given cluster image, all single-epoch images in the DES $r-$band which overlap with the central cluster galaxy (within 9$\arcmin$) are averaged to reduce variations in the sky background. The typical total exposure time for each cluster is 450 seconds, consisting of five 90 seconds single-epoch images from the first three years of DES observing. Because bright stars or nearby galaxies can affect diffuse light measurements, we remove clusters that are anywhere nearer than 526$\arcsec$ (equivalent to 2,000 pixels at DECam pixel scale) from these objects (using bad region mask > 2 described in \citealt{DrlicaWagner2018}). The single-epoch images of each cluster are then coadded together using the SWarp software \citep{2010ascl.soft10068B} to create one image for each cluster. The single-epoch images have been subtracted of sky background which are evaluated over the whole DECam field-of-view using a Principal Component Analysis (PCA) method \citep{Bernstein2017} for each exposure image, and the SWarp sky subtraction function is turned off during the coadding process.  

To isolate diffuse light from galaxies and foreground or background objects in the cluster field, we use the DES coadded object catalog to mask detected astronomical objects, but exclude the redMaPPer selected cluster central galaxies. The masks are constructed as ellipses with inclination, major and minor axes provided by the DES DM coadd catalog described in \citet{Abbot18DesDr1}. Figure ~\ref{fig:masking} shows examples of three \RM\ clusters ($z \sim 0.27$) analyzed in this paper before (top panel) and after masking (bottom panel). Unlike \citetalias{Zhang2019}, in which object brightness and detection significance cuts are applied before masking and then the faint galaxy contribution is estimated using the galaxy luminosity function constraints, we mask all objects to the DES Y3 catalog limit with detection S/N $ > 1.5$ (\texttt{magerr\_auto\_i} $ < 0.72 $). To avoid the presence of spurious objects we also apply a cut of \texttt{mag\_auto\_i} $ < 30.0 $, which is beyond the limiting magnitude of the Y3 catalog in $r$-band ($24.08$ mag for point-like sources at S/N = 10 within a 1.95$^{\prime\prime}$ diameter aperture as in \citealt{Abbot18DesDr1}). The improved Y3 catalog depth and our generous masking limit should eliminate any real objects detected in the images and we do not apply a faint galaxy contribution, as \citetalias{Zhang2019} demonstrated this component to be insignificant at redshift $\sim 0.25$ in DES data. 

For each masking object, the masking aperture is set to be 3.5 Kron radius of the object, 1.4 times as large in radius and 1.96 times as large in area as \citetalias{Zhang2019}. We hence avoid applying an unmasked residue light correction in the process -- the enlarged masking aperture would reduce the cluster galaxy residue contribution by $\sim$50\% compared to \citetalias{Zhang2019}. Calculation assuming Sérsic models states that a 3.5 Kron radius masking aperture only misses 0.8\% of the total light for a galaxy with Sérsic index $n=1$ (comparing to 4.2\% with a 2.5 Kron radius masking aperture). For $n = 4$ and $n = 8$, masking with 3.5 (2.5) Kron radius  misses 5.6\% (9.5\%) and 1.8\% (3.7\%) of the total light respectively. These fractions have been reduced by about 50\%, compared to using a 2.5 Kron radius masking aperture. In Z19, using a 2.5 Kron radius masking aperture and assuming a 9.6\% residue rate for all galaxies, the unmasked cluster galaxy light would make up $\sim$14\% of the measured diffuse light. We expect the ratio to be around 7\% using the 3.5 Kron radius apertures, but likely smaller than 7\%.

After the masking process, as mentioned in Sec.~\ref{sec:redmapperimgcat}, we further visually inspect all the clusters, and prune a total of 10 clusters that appear to be incompletely masked (because of image and catalog mismatching), or appear to be badly affected by nearby stars. 

The diffuse light profiles are then calculated as the average pixel values in the unmasked regions of the images in radial annuli, from which we then subtract residual background profiles to acquire the final measurements. 
The residual background profiles are measured around redMaPPer random points, which uniformly sample the sky coverage of the redMaPPer clusters in DES data. The same measurement process applied to the redMaPPer central galaxies, including masking and averaging pixel values in circular annuli, are applied to the random points. Thus we expect the residual background measurements to contain fluxes of sky background residuals as well as fluxes from undetected foreground and background astronomical sources \citep{2020arXiv200405618E}. We do require the random points to be at least 5 arcmin away from the cluster centers to avoid over-subtraction and a total number of 3859 random points are used in our analysis.
 
In the further measurement process, the clusters and random points are assigned to 40 regions using the \textit{Kmeans} code\footnote{\url{https://github.com/esheldon/kmeans\_radec}} \citep{Steinhaus1956}, which uses a clustering algorithm to divide the sky coverage of the redMaPPer clusters into regions with approximately the same area. We average the random point radial profiles in each region and use it as an estimation of the sky background of that region. This averaged random profile is subtracted from each of the measured cluster radial profiles in the same region. Each of the subtracted cluster profiles is then corrected to an observer frame at redshift z = 0.275 (median redshift of the sample), accounting for both distance dimming and angular-to-physical distance conversion. Finally, we sample the averaged cluster profiles using the jackknife method to estimate their uncertainties. Differently from \citetalias{Zhang2019}, we do not subtract the average flux value in the last radial bin of the image. 

In \citetalias{Zhang2019}, this measurement process has been tested by stacking random points and simulated diffuse light profiles which shows that the random background subtraction and the averaging process produce bias-free measurements. Discussions have also been undertaken about the influence of sky background estimations and the effect of instrument point spread function (PSF) on diffuse light interpretations. We refer the readers to that paper for further details regarding the measurement methods and tests.

\subsection{Surface brightness in luptitude}
\label{sec:lupprof}

While it is traditional to quantify diffuse light surface brightness in the unit of mag/area, for sky subtracted low surface brightness measurements near the noise limit which can be negative in flux, this leads to extremely noisy figures which are hard to interpret. In this paper, we present the surface brightness measurements of diffuse light in terms of asinh magnitudes proposed by \citet{Lupton1999}, which is informally known as ``luptitudes" (and we use lup as a symbol for this unit).
The luptitude system behaves very closely to the traditional log-based magnitude in the high signal-to-noise (S/N) regime, but has the advantage of robustness in the low signal-to-noise (S/N) regime or even when the flux is negative.

We calculate luptitude and its uncertainty from diffuse light flux and uncertainty using the following equations,
\begin{equation}
\begin{split}
\mu &= m_0 - 2.5 \log_{10} b - a \times \sinh^{-1} \left(\frac{f}{2b}\right), \text{ and} \\
\sigma_{\mu} & = \sqrt{ \frac{a^{2} \sigma_f^{2}} {4 b^{2} + f}}.
\label{eq:luptitude}
\end{split}
\end{equation}
In these equations, $m_0$ = 30 is the zeropoint magnitude; $a$ $\equiv$ 2.5 $\log_{10}\mathrm{e}$ = 1.0857 (Pogson ratio); $\sigma_f$ is the flux ($f$) measurement uncertainty; $b$ is the softening parameter, or knee of the luptitude function where standard magnitudes and luptitudes begin to significantly diverge, and is defined as $b\equiv \sqrt{a} \sigma_{f} \equiv 1.042 \sigma_{f}$, in which $\sigma_{f}$ is fixed to be a flux uncertainty at 500 kpc which sets $b=0.66571579$.

\section{Diffuse light profiles}
\label{sec:massdep}

\subsection{Flux profile and integrated flux }
\label{sec:fluxprof}

\begin{figure}
	\includegraphics[width=\columnwidth]{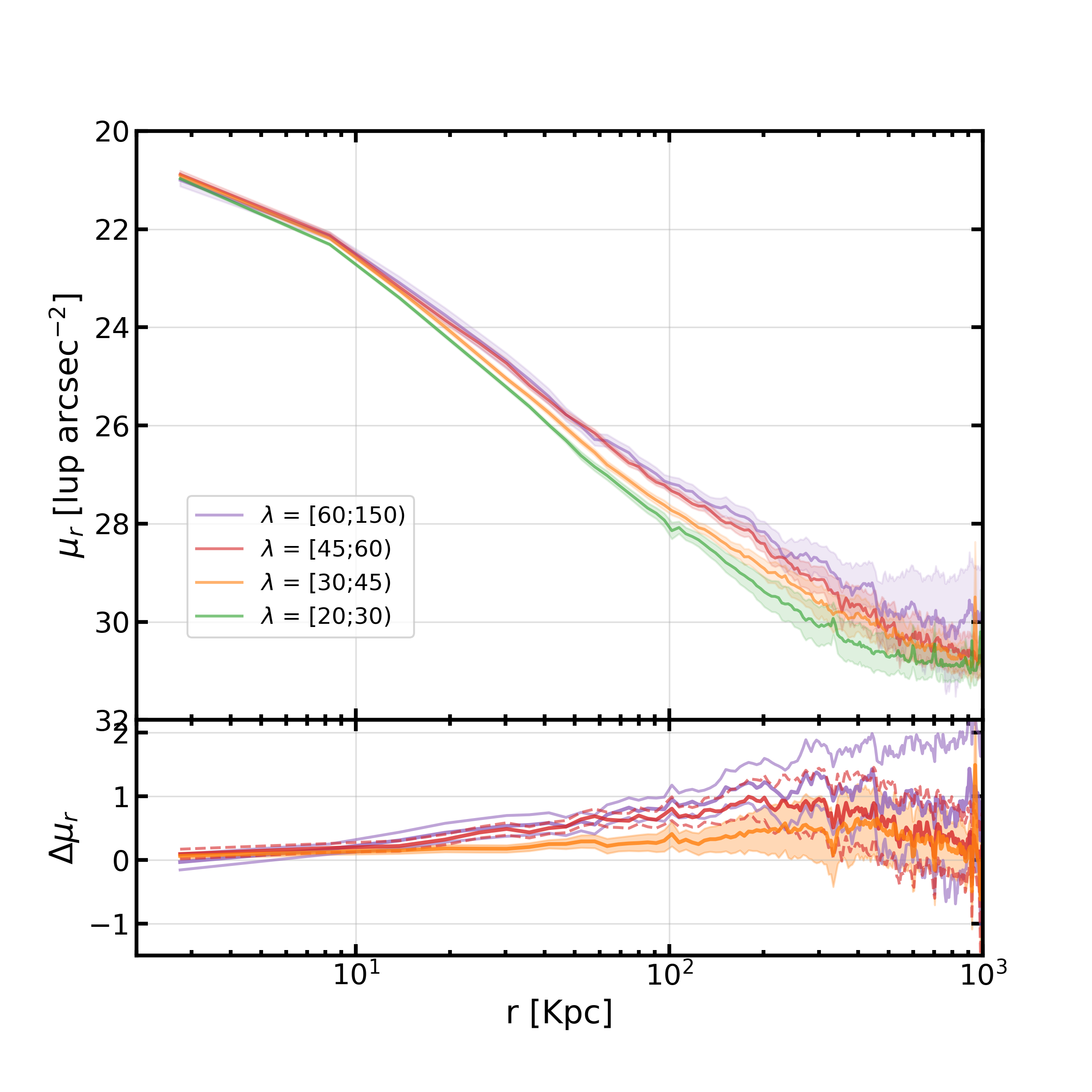}
    \caption{{\bf Top panel:} Stacked surface brightness profiles of clusters in different richness bins. The shaded regions represent the uncertainties, which are computed through jackknife sampling. The diffuse light profiles show similar profiles in the center regions, but more massive clusters have a higher level of surface brightness in the outskirts (mass dependence). {\bf Bottom panel:} The difference between the lowest richness bin profile, used as a reference, and other richness bins profiles. For reference, at redshift 0.275, 1 arcsec $\approx$ 4.2 kpc.}
    \label{fig:iclprofiles_lup}
\end{figure}

\begin{figure}
	\includegraphics[width=\columnwidth]{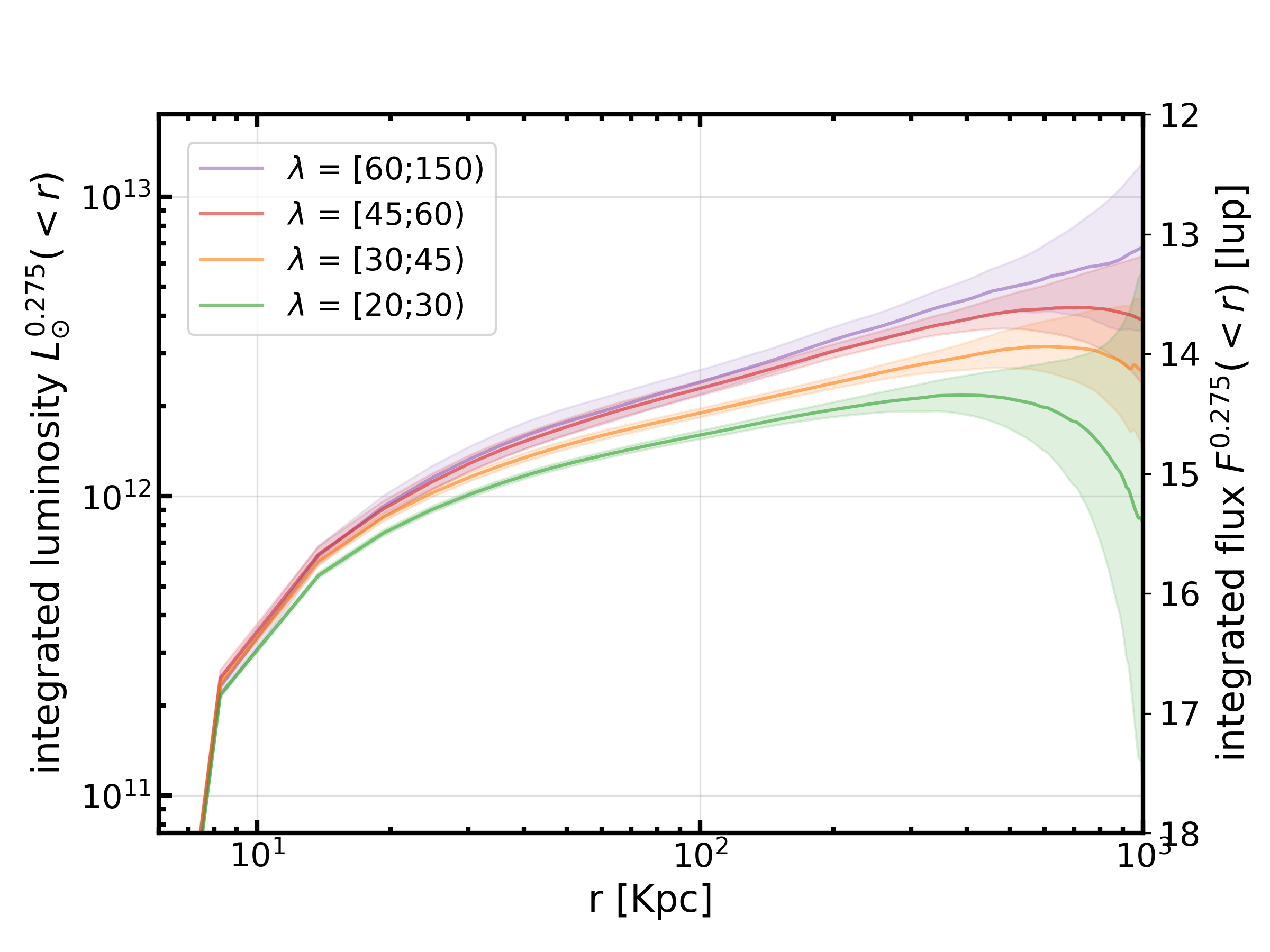}
    \caption{Integrated diffuse light profiles of clusters in different richness bins in luptitude (left $y-$axis) and solar luminosity unit (right $y-$axis). The shaded regions represent the uncertainties which are computed through jackknife sampling. The integrated diffuse light profiles show similar profiles in the center regions, but more massive clusters have a higher level of integrated surface brightness in the outskirts (mass dependence). Note that because diffuse light flux may fluctuate to negative values because of sky subtraction, the integrated flux may show features that decrease with radius in low S/N regions.}
    \label{fig:int_mag_lup}
\end{figure}

As mentioned in Sec. \ref{sec:redmapperimgcat}, we divide the clusters into 4 richness sub-samples following \citet{McClintock2019}, 20 $\leq$ $\lambda$ < 30, 30 $\leq$ $\lambda$ < 45, 45 $\leq$ $\lambda$ < 60 and 60 $\leq$ $\lambda$ < 150, which correspond to mean masses of 1.6 $\times$ 10$^{14}$, 2.7 $\times$ 10$^{14}$, 4.3 $\times$ 10$^{14}$ and 8.0 $\times$ 10$^{14}$ M$_\odot$. 
We compute the diffuse light surface brightness profiles as described in Sec. \ref{sec:data}, accounting for both distance dimming and angular-to-proper distance (at redshift 0.275, 1 arcsec $\approx$ 4.2 kpc) and convert the fluxes to luptitudes with a zero-point of 30 (see Sec. \ref{sec:lupprof}). These surface brightness profiles are also integrated to derive the total diffuse light luminosity as a function of radius, as in
\begin{equation}
  F(R) = 2\upi\int_{0}^{R} r' L(r') dr'
  \label{eq:integrated}
\end{equation}
where $L(r')$ is the flux profile. Figures \ref{fig:iclprofiles_lup} and \ref{fig:int_mag_lup} show, respectively, the surface brightness and integrated brightness profiles of cluster diffuse light in different richness ranges. We estimated the $1 ~\sigma$ background fluctuations in the radial range of 0.8 Mpc to 1 Mpc to be 0.451, 0.450, 0.498 and 1.394  flux per arcsec$^{2}$ (magnitude zeropoint is 30) in the four cluster richness bins considered in our study, which correspond to surface brightness limits of 30.08, 30.08, 30.04 and 29.45 lup/arcsec$^{2}$.

Unsurprisingly, the surface brightness and integrated brightness of diffuse light in richer clusters is brighter, which can be explained given that richer and thus more massive clusters host more satellite galaxies \citep{Gao2004}, and tidal stripping as well as dwarf galaxy disruption have the opportunity to disperse more stars into the intracluster space. However, the surface brightness and integrated brightness of diffuse light in the cluster central region varies little with cluster richness. This effect is in agreement with the inside-out growth scenario, which assumes that galaxy centers form early in a single star-burst, and the accreted galaxy stellar content at later times are deposited onto the galaxy outskirts (e.g. \citealt{vanDokkum2010}; \citealt{vanderBurg2015}; \citealt{2010ApJ...725.2312O}). These effects have also been noted in \citetalias{Zibetti2005} and \citetalias{Zhang2019}.

\begin{figure}
	\includegraphics[width=\columnwidth]{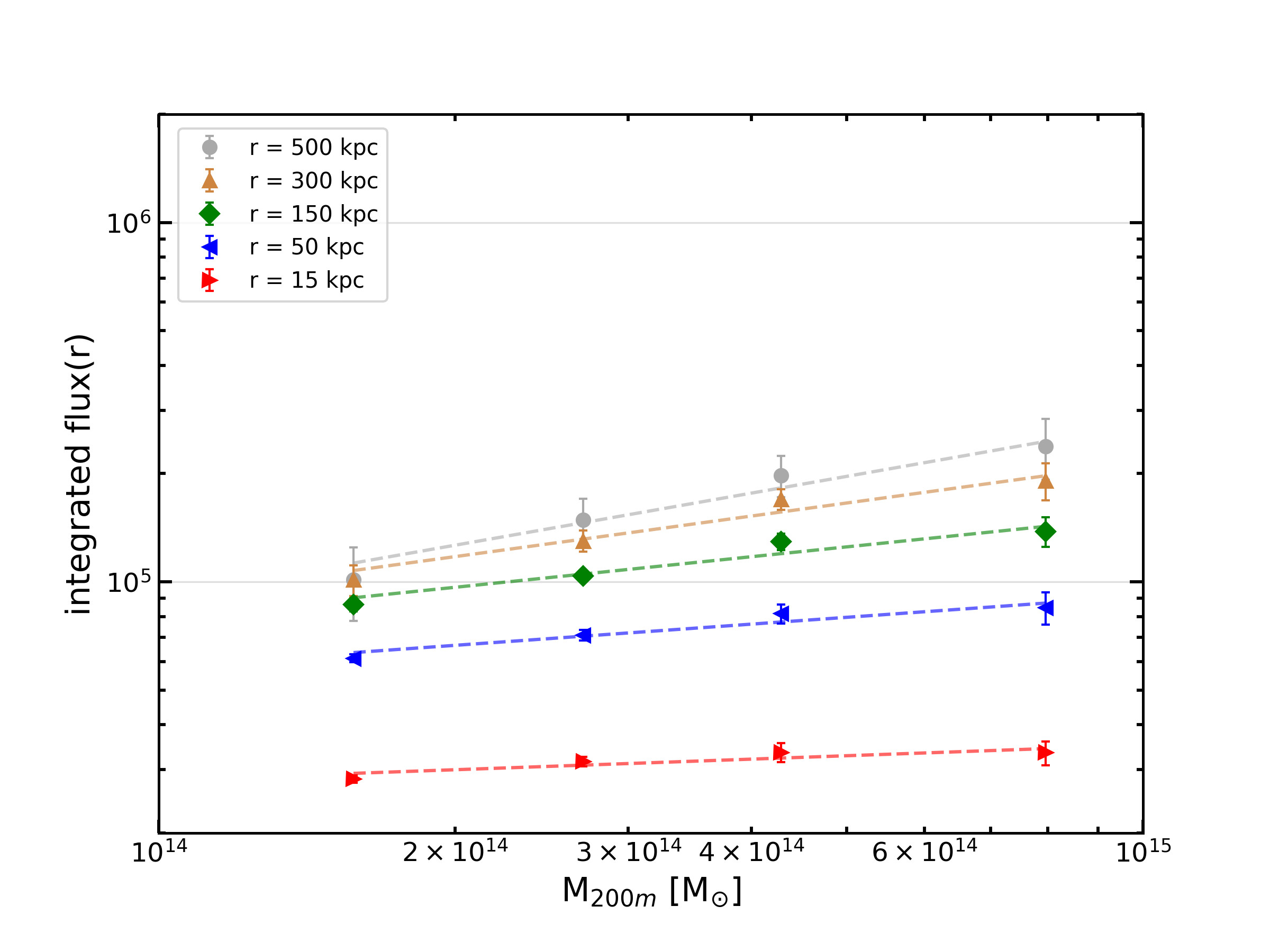}
    \caption{Cluster mass dependence of the integrated diffuse light flux. We compute the integrated fluxes within 5 radii (15, 50, 150, 300 and 500 kpc) and show them as a function of the cluster mass. The dotted lines show a best linear fitting to the logarithmic values  of the flux and the cluster mass. The mass dependence slope  becomes steeper with larger radii, indicating stronger correlation between diffuse light luminosity and cluster mass.}
    \label{fig:rrdep}
\end{figure}

We further investigate the mass dependence of the diffuse light integrated fluxes within five radii, 15, 50, 150, 300, and 500 kpc, which range from being dominated by the BCG, to being dominated by the diffuse light. We use the cluster mass estimations modeled from cluster weak lensing measurements in \cite{McClintock2019}. Figure \ref{fig:rrdep} show the integrated fluxes in these radial ranges as a function of the cluster mass. 
To examine the steepness of the cluster mass dependence, we perform a linear fit to the logarithmic values of the integrated diffuse flux versus $M_{200\text{m}}$, as
\begin{equation}
    \log_{10} F(R) = \alpha \log_{10} M_{200\text{m}} + \beta,
	\label{eq:linear_fit}
\end{equation}
where $\alpha$ is the slope and $\beta$ is the y-intercept. We also estimate the Pearson correlation coefficient ($\rho_{\text{cc}}$) as,
\begin{equation}
    \rho_{\text{cc}} = \frac{\mathrm{Cov}(\log_{10} M_{200\text{m}}, \log_{10}F(R))}{\sqrt{\mathrm{Var}(\log_{10} M_{200\text{m}} ) \textrm{Var}(\log_{10}F(R))}},
	\label{eq:pearson}
\end{equation}

We report the best-fit parameter values and the correlation coefficients in Tab. \ref{table:lfit1}. The slope of the flux-$M_{200\text{m}}$ dependence is insignificant at small radii (15 and 50 kpc), but becomes steeper with enlarging radius and is most pronounced at the largest radius. The correlation between total diffuse light luminosity and cluster mass is excellent at large radius beyond 50 kpc: the fitting slope indicating the diffuse light mass-dependence is steep and significant at 500 kpc; the correlation coefficient values is also significant, reaching $\rho_{\text{cc}} > 0.9$ outside of 300 kpc. We will return to this correlation and further explore the connection between diffuse light and cluster masses in the upcoming sections.

\begin{table}
\centering 
\caption{Best linear-fit parameters for integrated profiles.} 
\begin{tabular}{cccc}
\hline
\multicolumn{4}{c}{Diffuse Light}  \\
\hline
Radius [kpc] & $\alpha$ & $\beta$ & $\rho_{\text{cc}}$\\
\hline\hline 
15 & 0.103$\pm$0.049 & 4.328$\pm$0.084 & 0.719 \\ 
50 & 0.209$\pm$0.064 & 4.520$\pm$0.112 & 0.847 \\ 
150 & 0.300$\pm$0.088 & 4.550$\pm$0.155 & 0.870 \\ 
300 & 0.397$\pm$0.102 & 4.495$\pm$0.183 & 0.900 \\ 
500 & 0.507$\pm$0.118 & 4.368$\pm$0.215 & 0.922 \\ 
\hline
\multicolumn{4}{c}{Cluster Total Light}  \\
\hline
Radius [kpc]    & $\alpha$ & $\beta$ & $\rho_{\text{cc}}$ \\
\hline
\hline
15 & 0.133$\pm$0.049 & 2.582$\pm$0.721 & 0.727 \\ 
50 & 0.247$\pm$0.013 & 1.327$\pm$0.186 & 0.948 \\ 
150 & 0.203$\pm$0.050 & 2.354$\pm$0.724 & 0.864 \\ 
300 & 0.349$\pm$0.066 & 0.471$\pm$0.972 & 0.918 \\ 
500 & 0.442$\pm$0.059 & -0.709$\pm$0.861 & 0.983 \\ 
\hline
\end{tabular}
\label{table:lfit1} 
\end{table}

Note that the above measurements are made around the redMaPPer-selected central galaxies, which aim to select the cluster galaxies closest to the peak of the cluster matter distribution. Studies have found these selections to be correct for $\sim75\pm 8 \%$ of the clusters \citep{Zhang2019Misc}, but the redMaPPer algorithm may misidentify a cluster satellite galaxy, or a projected foreground/background galaxy as the center \citep{Hollowood2019}. 
Here, we briefly estimate the effect of miscentering using a formula often adopted in cluster weak lensing analyses \citep[e.g.][]{McClintock2019}, where a well-centered cluster model is integrated over a miscentering offset distribution, weighted by the fraction of miscentered clusters. We model the diffuse light in well-centered clusters as the sum of two Sérsic components, one with a half-light radius R$_e$ of 52.1 kpc and another with a R$_e$ of 2.6 Mpc, as specified in \citetalias{Zhang2019}. In addition, the miscentered and well-centered clusters, all have a core central galaxy component with a R$_e$ of 9.13 kpc, since redMaPPer always pick a galaxy as the center, even if not necessarily the right one. 
The miscentering offset shifts the measurement of diffuse light in the inner part of the clusters to outer parts. Overall, it increases the measured flux of diffuse light surface brightness by 10 to 20\% beyond 200 kpc radius, while reduces it within the radius range of 100 to 200 kpc.

However, miscentering should have minimal effects on the rest of our results when comparing diffuse light, total cluster light, and cluster weak lensing measurements, since those are measured around the same central galaxies.

\subsection{Self-Similarity}
\label{sec:self_similarity}

The distribution of dark matter, hot gas, and even member galaxies in galaxy clusters are known to exhibit a large degree of self-similarity, so that these cluster components follow a nearly universal radial profile after scaling by a characteristic radius related to the cluster's mass and redshift (e.g., dark matter: \citealt{1997ApJ...490..493N}, hot gas: \citealt{1986MNRAS.222..323K}, cluster galaxies: \citealt{2012MNRAS.423..104B}). These extraordinary properties often mean a low scatter relation that relates the cluster's dark matter, gaseous or satellite galaxy observables to the cluster's total mass.

In \citetalias{Zhang2019}, it was discovered that cluster diffuse light also appears to be self-similar, i.e., clusters of different masses appear to have a universal diffuse light profile at large radii beyond 100 kpc of the cluster center, after scaling by the cluster's $R_{200\text{m}}$, indicating a tight relation between diffuse light and cluster mass. In this section, we revisit the diffuse light self-similarity by scaling the surface brightness profiles by $R_{200\text{m}}$. For each cluster richness sub-sample, we estimate their $\langle R_{200\text{m}} \rangle$ using,
\begin{equation}
    \langle R_{200\text{m}}\rangle = \sqrt[3]{\frac{3 \langle M_{200\text{m}}\rangle}{800\upi\rho_{\text{m}}(z_{\text{m}})}},
	\label{eq:r200m}
\end{equation}
where $M_{200\text{m}}$ is the mean mass of each sub-sample estimated with the mass-richness relation from \citet{McClintock2019} and $z_{\text{m}}$ is the mean cluster redshift, 0.275; $\rho_{\text{m}}(z_{\text{m}})$ = $\Omega_{\text{m}}$ $\rho_{\text{crit}}$ (1+$z_{\text{m}}$)$^{3}$ is the mean cosmic matter density in physical units for $z_{\text{m}}$, $\rho_{\text{crit}}$ is the critical density at
redshift zero.
The $\langle R_{200\text{m}} \rangle$ values are estimated to be 1305.76, 1561.73, 1822.72, 2240.30 kpc at 20 $\leq$ $\lambda$ < 30, 30 $\leq$ $\lambda$ < 45, 45 $\leq$ $\lambda$ < 60 and 60 $\leq$ $\lambda$ < 150, respectively.

\begin{figure}
	\includegraphics[width=\columnwidth]{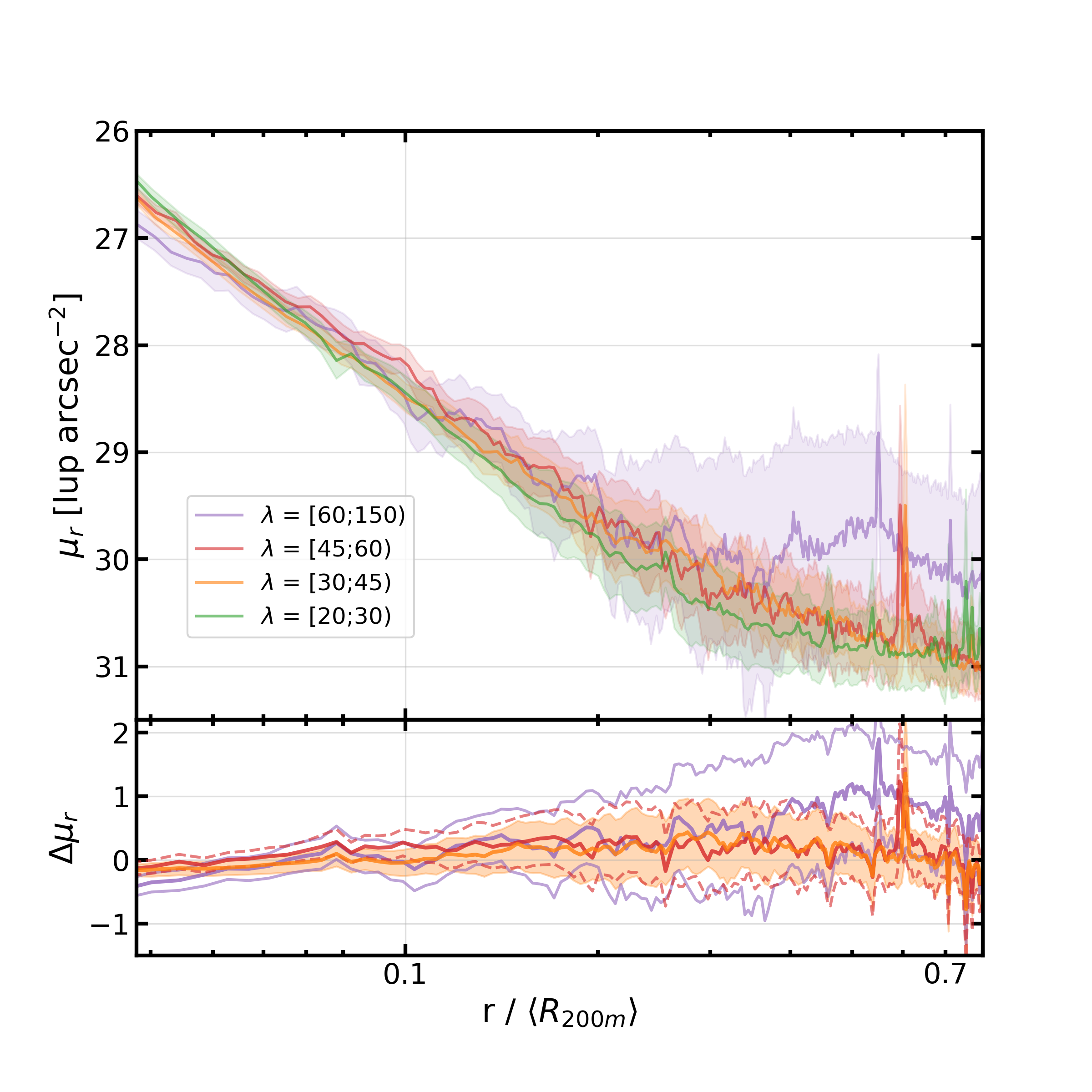}
    \caption{Scaled diffuse light profiles for different cluster richness bins. The smallest radii in this figure would correspond to 52.23, 62.47, 72.91 and 89.61 kpc before scaling the profiles by cluster $\langle R_{200\text{m}} \rangle$ for richnesses 20 $\leq$ $\lambda$ < 30, 30 $\leq$ $\lambda$ < 45, 45 $\leq$ $\lambda$ < 60 and 60 $\leq$ $\lambda$ < 150, respectively; while the largest radii would correspond to 1044.61, 1249.39, 1458.18 and 1792.24 kpc for richnesses 20 $\leq$ $\lambda$ < 30, 30 $\leq$ $\lambda$ < 45, 45 $\leq$ $\lambda$ < 60 and 60 $\leq$ $\lambda$ < 150, respectively. The profile uncertainties are represented by the shaded regions and estimated using jackknife sampling. All profiles show self-similarities up to 0.8 r/$R_{\mathrm{200m}}$ within 1$\sigma$. {\bf Bottom panel:} The difference between the lowest richness bin profile, used as a reference, and other richness bins profiles.}
    \label{fig:self_sim}
\end{figure}

Figure  \ref{fig:self_sim} shows the diffuse light profiles after scaling by $\langle R_{200\text{m}} \rangle$. We observe self-similarity between all the richness bins outside 0.05 r/$R_{\mathrm{200m}}$ and up to 0.8 r/$R_{\mathrm{200m}}$ within 1$\sigma$.

\subsection{Cluster total light}
\label{sec:tot_l_prof}

We also derive the radial profiles from the cluster images without masking any objects (as shown in the top panels of Figure  \ref{fig:masking}). When none of the objects are masked, the cluster images not only contain the light from the diffuse light, but also the rest of the cluster galaxies. The images also contain light from the foreground and background structures, although these contributions are eliminated later by subtracting light profiles derived from random images.  Throughout this paper, we refer to the light profiles derived from the unmasked images as the cluster total light profiles. 

For the computation of these cluster total light profiles, we follow the same procedure as described in Sec. \ref{sec:lprof}, with the exception that we use the unmasked images for both clusters and random points. When computing the sky brightness level using the unmasked random images, the sky brightness level obtained is higher than that from the masked random images, because we are observing the contribution of all the components of the image. We apply the subtraction between the unmasked cluster images and the random images to derive the cluster total light profiles. We notice these radial profiles to be much noisier at radii larger than r = 25 kpc, compared to the diffuse light profiles. Thus, for the regions beyond 25 kpc, we use coarser radial bins to improve the signal-to-noise. We use 15 radii bins in logarithmic space beyond 25 kpc. The uncertainties of the cluster total light profiles are sampled with the jackknife method applied to the individual profiles.

\begin{figure}
	\includegraphics[width=\columnwidth]{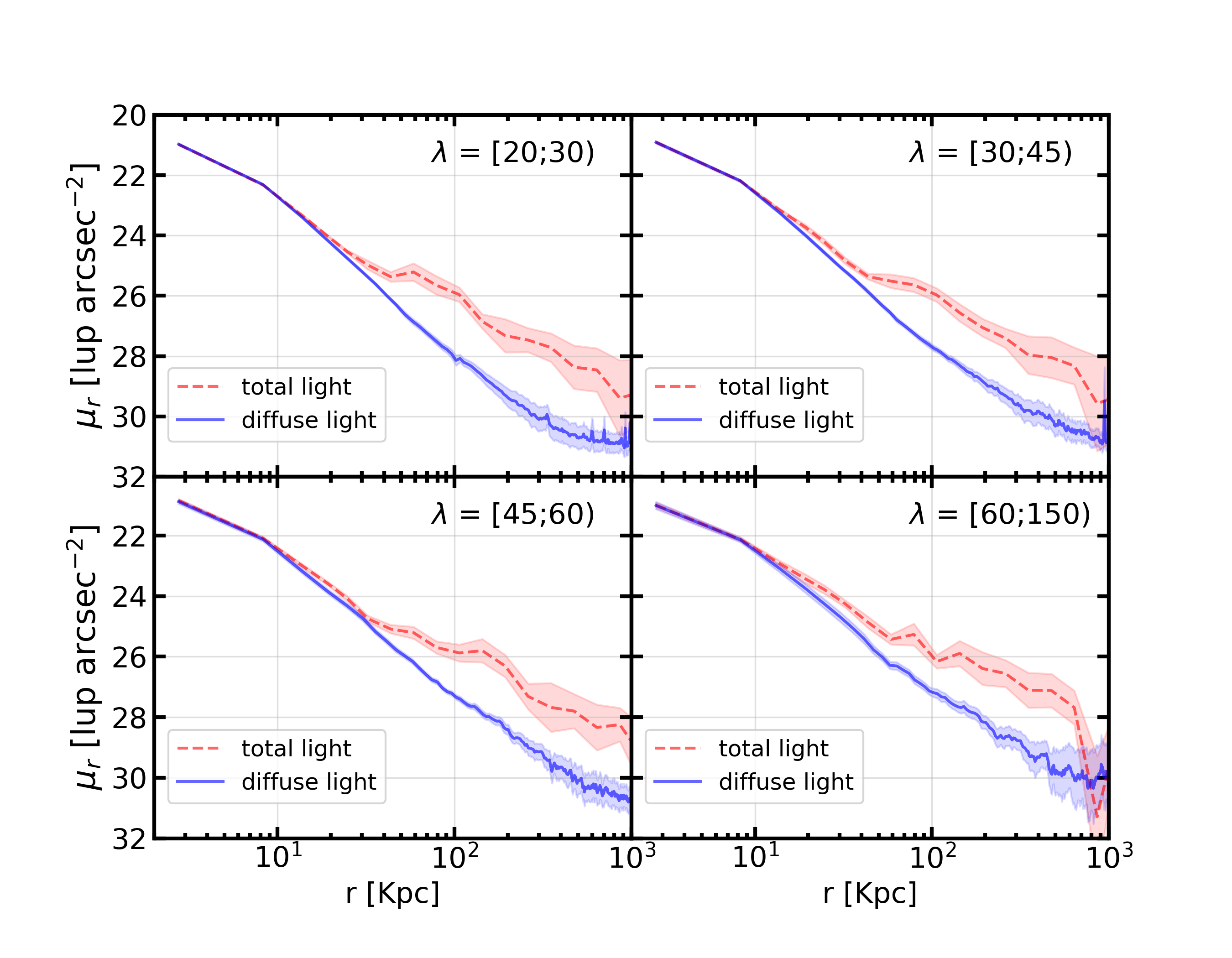}
	\includegraphics[width=\columnwidth]{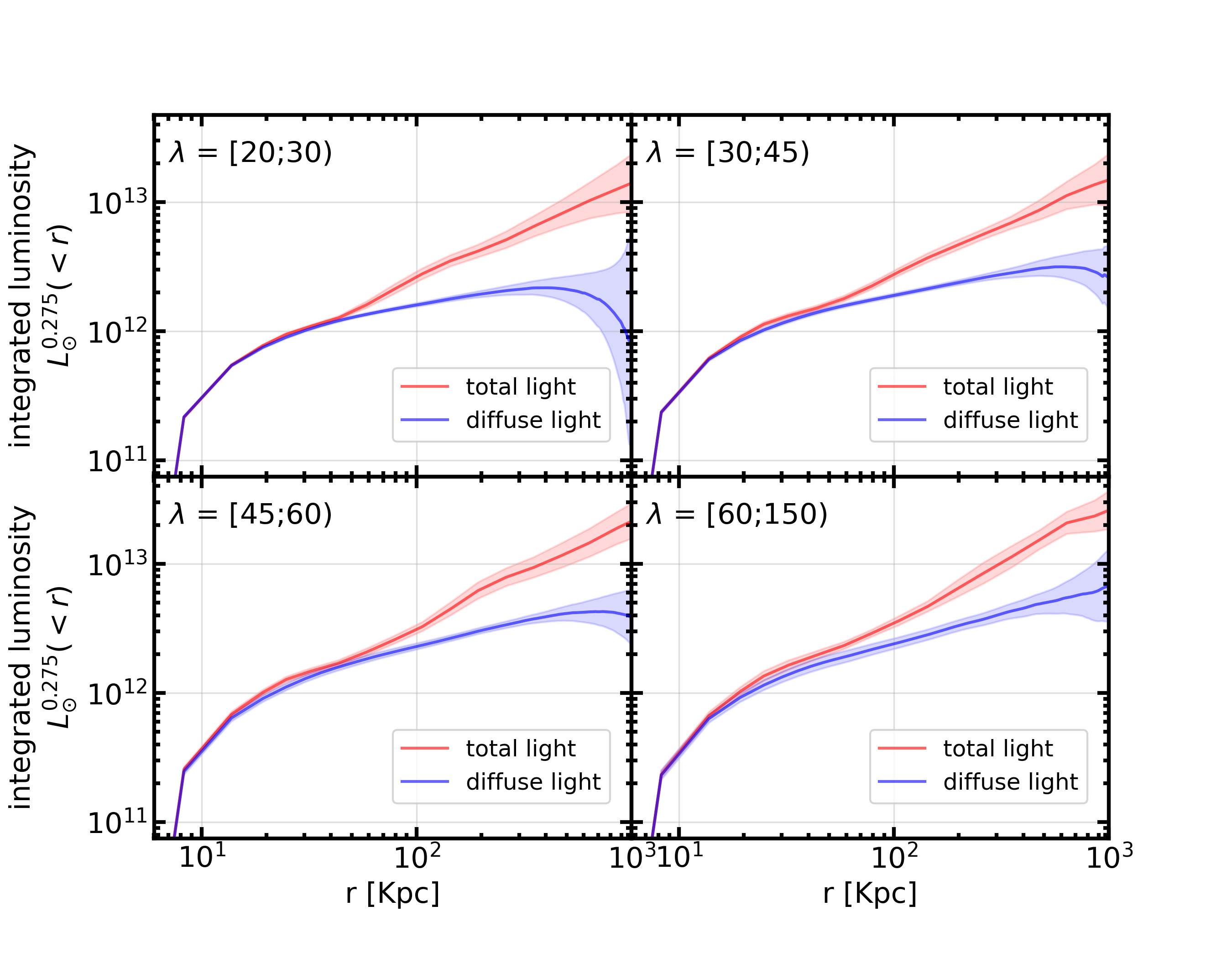}
    \caption{Cluster total light (red dashed line) and the diffuse light (blue solid line) profiles in different cluster richness bins, in terms of surface brightness (\textbf{upper panels}) and integrated fluxes (\textbf{lower panels}) measured in the observer frame at redshift 0.275. The profile uncertainties are represented by the shaded regions and estimated using jackknife sampling. For reference, at redshift 0.275, 1 arcsec $\approx$ 4.2 kpc.}
    \label{fig:mag_unm_lup_comparison}
\end{figure}

\begin{figure}
	\includegraphics[width=\columnwidth]{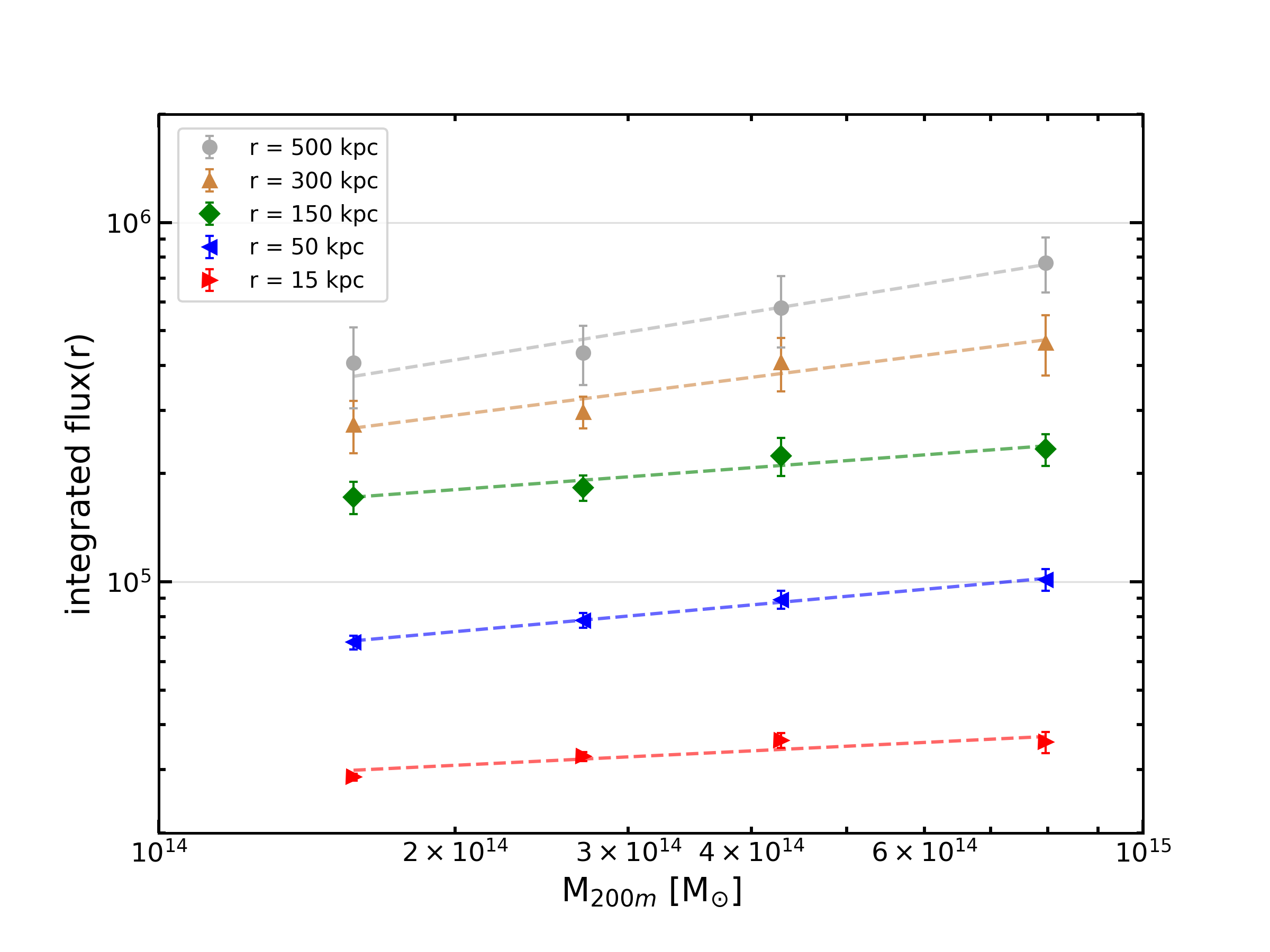}
    \caption{Cluster mass dependence of the integrated cluster total light flux. We compute the integrated fluxes within 5 radii (15, 50, 150, 300 and 500 kpc) and show them as a function of the cluster mass. The dotted lines show the best linear fitting results to the logarithmic values of the flux and the cluster mass. The slope of the linear fits become steeper with larger radii.}
    \label{fig:rrdep_unmp}
\end{figure}

Figure  \ref{fig:mag_unm_lup_comparison} displays the cluster total light profiles in comparison to the diffuse light profiles. 
Both diffuse light and cluster total light profiles become fainter as the radius increases, with the total light surface brightness reaching $\sim$ 28 lup/arcsec$^{2}$ at r = 500 kpc. Since the cluster total light is completely dominated by the BCG light within r $\sim$ 10 kpc, the cluster total light and diffuse light profiles coincide in this radial range. The bottom panels of Figure  \ref{fig:mag_unm_lup_comparison} further show the integrated radial profiles of the diffuse light and total light. The total light in the richest clusters reaches a  brightness of 2.6 $\times$ 10$^{13}$ $L_{\odot}$ at r = 1 Mpc, and the cluster total light deviates significantly from the diffuse light beyond $\sim$ 100 kpc.

As in Sec.~\ref{sec:fluxprof}, we derive the integrated flux of cluster total light in five radial ranges, and study their mass dependence as shown in Figure  \ref{fig:rrdep_unmp}. A linear fit to the logarithmic values between the integrated flux and the cluster mass, $M_{200\text{m}}$, is performed and the best-fit parameters are reported in the lower section of Tab. \ref{table:lfit1}. The diffuse light and cluster total light flux show increasing mass dependence at larger radius and are well correlated with cluster total mass. At 500 kpc, the diffuse light and cluster total light flux both have significant correlation coefficient values with $\rho_{\text{cc}} > 0.9$, and steep mass dependence slopes of 0.507$\pm$0.118 and 0.442$\pm$0.059 respectively.

\subsection{Diffuse light to cluster total light fraction}
\label{sec:ratio}

\begin{figure*}
	\includegraphics[width=2\columnwidth]{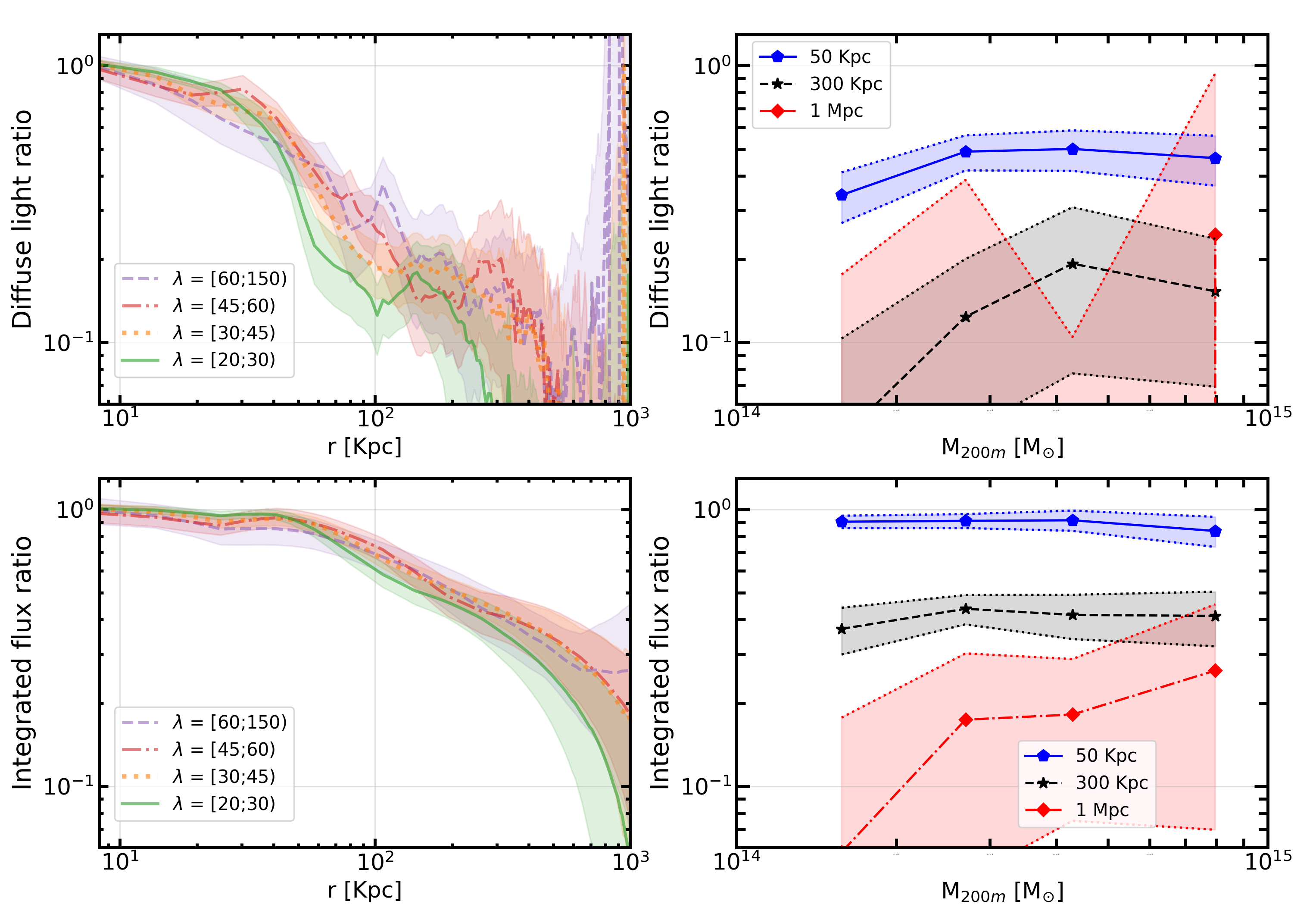}
    \caption{\textbf{Upper panels:} Diffuse light fraction in the cluster total light, as a function of radius (left panel) and the cluster total mass measured. \textbf{Lower panels:} Cumulative diffuse light fraction in the cluster total light, as a function of radius (left panel) and cluster total mass (right panel). The mass dependence increases mildly with radius, whereas the diffuse light ratio at 50 kpc presents no trend with increasing of mass and a mild trend at 300 kpc; while integrated flux ratio presents no trend at 50 kpc and 300 kpc and a mild trend at 1000 kpc.}
    \label{fig:icl_frac}
\end{figure*}

A very important property of diffuse light is its fraction in total cluster light. 
We  measure this property by dividing the surface brightness profiles and the integrated profiles of the diffuse light by the corresponding profiles of total cluster light,  
\begin{equation}
    \begin{split}
    f_{\mathrm{frac}}(R) & = \frac{L_{\mathrm{diffuse}}(R)}{ L_{\mathrm{total}}(R)}, \\
    F_{\mathrm{frac}}(R) & = \frac{F_{\mathrm{diffuse}}(R)}{ F_{\mathrm{total}}(R)},
    \end{split}
    \label{eq:ratio}
\end{equation}
which is the diffuse light fraction and cumulative diffuse light fraction respectively.
Figure  \ref{fig:icl_frac} shows these fractions in different cluster radial and richness ranges. 
We report those fractions at 50, 300, 700 and 1000 kpc for 20 $\leq$ $\lambda$ < 30, 30 $\leq$ $\lambda$ < 45, 45 $\leq$ $\lambda$ < 60 and 60 $\leq$ $\lambda$ < 150 in Tab. \ref{table:frac}. Within 50 kpc, diffuse light makes up most of the cluster total light, and the cumulative fraction is above 80\% regardless of cluster richness. Beyond 50 kpc, given the faster increase of cluster total light with radius than diffuse light, the cumulative diffuse light fraction steadily decreases with increasing radius, which reaches around $\sim$24\% at 700 kpc regardless of cluster richness. We do not notice obvious cluster richness/mass dependence of the diffuse light fractions, especially beyond 200 kpc.

\begin{table}
\centering
\caption{The diffuse light surface brightness fraction and cumulative flux fraction in cluster total light at various cluster radii. } 
\begin{tabular}{c|cccc}
\hline
 & \multicolumn{4}{c}{Surface Brightness fraction (\%)}  \\
\hline
$\lambda$  & 50 kpc & 300 kpc & 700 kpc & 1 Mpc \\
\hline
\hline 
20-30 & 34.1$\pm$7.1 & 4.5$\pm$5.8 & -- & -- \\ 
30-45 & 48.9$\pm$7.1 & 12.4$\pm$7.7 & -- & -- \\ 
45-60 & 50.0$\pm$8.3 & 19.3$\pm$11.5 & -- & -- \\ 
60-150 & 46.3$\pm$9.5 & 15.3$\pm$8.4 & -- & -- \\  
\hline
& \multicolumn{4}{c}{Cumulative flux Fraction (\%)}  \\
\hline
$\lambda$ & 50 kpc & 300 kpc & 700 kpc & 1 Mpc \\
\hline
\hline
20-30 & 90.3$\pm$4.6 & 37.0$\pm$7.1 & 16.2$\pm$9.9 & 5.8$\pm$11.9 \\ 
30-45 & 91.0$\pm$5.3 & 43.8$\pm$5.3 & 26.2$\pm$9.9 & 17.4$\pm$12.8 \\ 
45-60 & 91.4$\pm$7.7 & 41.6$\pm$7.6 & 27.0$\pm$9.6 & 18.2$\pm$10.7 \\ 
60-150 & 83.6$\pm$10.4 & 41.3$\pm$9.2 & 26.3$\pm$10.7 & 26.2$\pm$19.2 \\ 
\hline
\end{tabular}
\label{table:frac} 
\end{table}

Many previous studies have measured diffuse light fraction, but the results seem to be at tension possibly caused by different analysis choices. An important consideration is that the diffuse light  fraction changes with the analysis radius, as our measurements demonstrate. Previously, \cite{Krick2007} found that the diffuse light fraction is between 6$\pm$5\% and 22$\pm$12\% at one-quarter of the virial radius using $r-$band, while \cite{Montes2018} found this fraction to be between 8.6$\pm$5.6\% and 13.1$\pm$2.8\% at R$_{500}$, and \citetalias{Zhang2019} measured a diffuse light fraction of 44$\pm$17\% at 1 Mpc. Our results of diffuse light fraction being $\sim$ 24\% at 700 kpc agrees with the ranges reported in the previous work. 

How the diffuse light fraction changes with cluster mass is another interesting topic in diffuse light studies. Efforts with semi-analytical studies have suggested an increasing diffuse light fraction with cluster mass, reaching around 50\% in clusters of 1.42 $\times$ 10$^{15}$ M$_{\odot}$ mass \citep{Lin2004}. Observationally \citet{Zibetti2005} found no evidence of mass dependence of the diffuse light fraction. 
Figure  \ref{fig:icl_frac} shows our results demonstrating the mass (in)dependence of the diffuse light fraction in three radii. There is no outstanding difference in the diffuse light fractions between cluster richness subsets within 300 kpc, which is in agreement with \citet{Zibetti2005}. However, since our results are derived in physical radius,  the diffuse light fractions will likely change with cluster mass when derived in terms of the normalized cluster radius such $R_\mathrm{200m}$. In addition, at large radius, we notice a low significance increase of the diffuse light fraction with mass, although higher signal-to-noise measurements will be needed to confirm this trend. 

\section{Comparison to weak lensing}
\label{sec:goodtracer}

Recent studies have presented significant evidence of a connection between diffuse light and the cluster dark matter (or total mass) distribution -- diffuse light profiles have similar radial slopes with the total cluster dark matter density distribution \citep[e.g.][]{Pillepich2014, Pillepich2018, Montes2018}; The diffuse light surface brightness contours are highly similar to the cluster mass density contours \citep[e.g.][]{Montes2019, Asensio2020}. In \citetalias{Zhang2019} and this paper (Figure~\ref{fig:self_sim}), we also note the diffuse light surface brightness to be self-similar, appearing to have a universal radial profile after scaling by cluster  $R_{200\text{m}}$ radius. 

These analyses raise an interesting question -- does diffuse light trace the cluster dark matter, and thus trace the cluster total matter distribution? In this section, we explicitly explore this question by comparing the diffuse light radial dependency with that of the cluster total matter measured through weak lensing.

\subsection{Weak-lensing measurements}
\label{sec:wl-measurements}

\begin{figure}
	\includegraphics[width=\columnwidth]{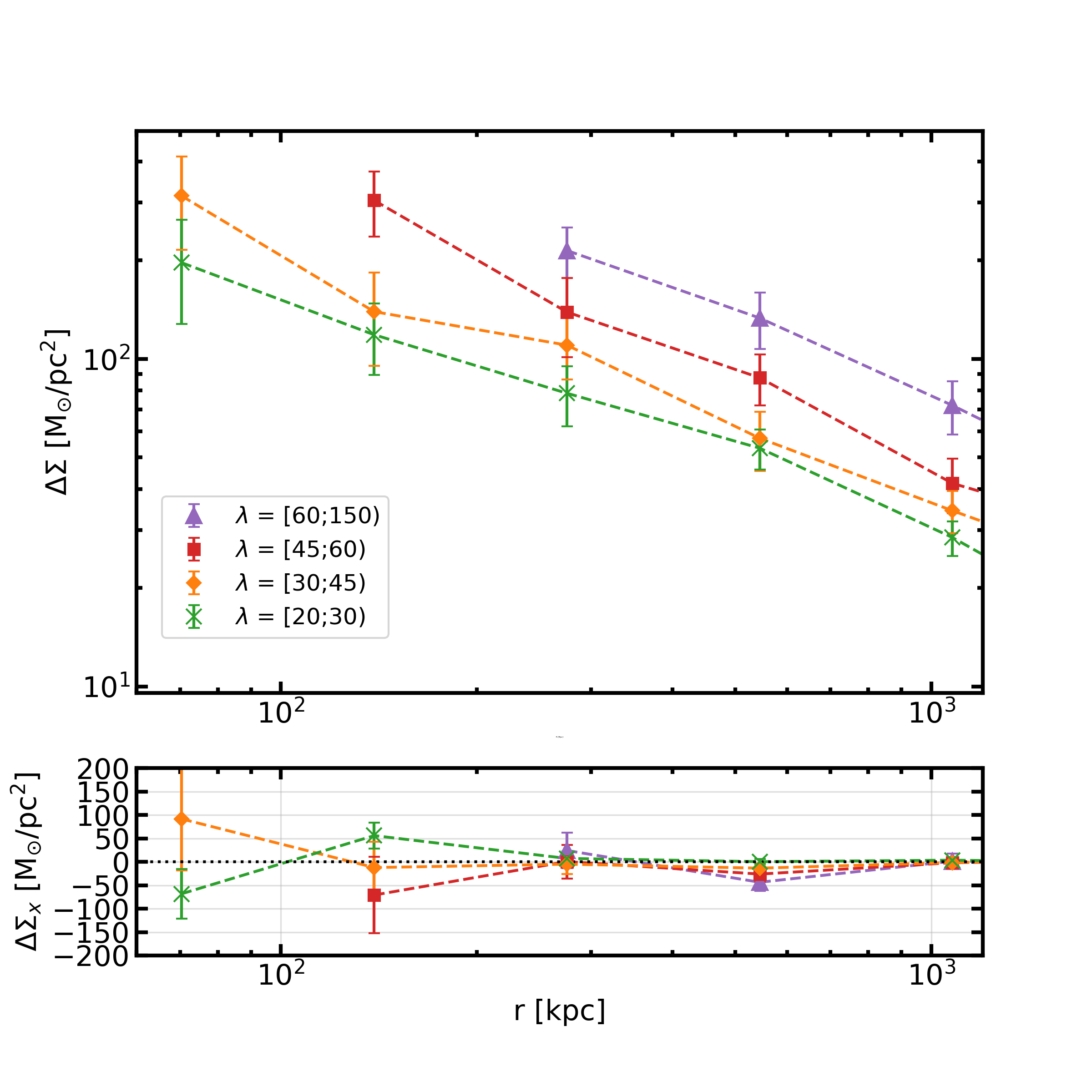}
    \caption{{\bf Upper Panel} shows the cluster mass profiles in four richness bins measured through WL tangential shear estimations. {\bf Lower panel} shows the validation cross-component shear measurements around the same clusters, which is consistent with a null signal and indicates the shear measurements to be relatively bias-free. The cluster mass profiles measured through WL tangential shear estimations is compared to the diffuse light profiles in Sec.~\ref{sec:goodtracer}.}
    \label{fig:wl}
\end{figure}

The cluster total matter radial distributions are derived through the tangential shear measurements from weak lensing  around the clusters of interest. The azimuthally averaged tangential shear is related to the 2-dimensional surface density as: 
\begin{equation}
    \gamma_{\rm T} = \frac{\bar{\Sigma}(<R)-\bar{\Sigma}(R)}{\Sigma_{\rm \text{crit}}}= \frac{\Delta\Sigma (R)}{\Sigma_{\rm \text{crit}}},
\end{equation}
where $\bar{\Sigma}(<R)$ is the average cluster surface mass density inside the radius of $R$, $\bar{\Sigma}(R)$ is the surface mass density at the radius of $R$, and $\Sigma_{\rm \text{crit}}$ is given as,
\begin{equation}
    \Sigma_{\rm \text{crit}}(z_{\rm s},z_{\rm l}) = \frac{c^2}{4 \upi G} \frac{D_{\rm s}}{D_{\rm l} D_{\rm ls}},
\end{equation}
where $z$ and $D$ denote the redshift and the distance to the object, respectively, and the subscripts s and l, the source and the lens. 
We use $\Delta\Sigma (R)$ for the following analyses, as described ahead.

The shape catalog \citep{Zuntz2018} used for the weak lensing measurements in this paper is produced by \textsc{Metacalibration} \citep{Sheldon2017}.
In contrast to other shear estimation algorithms, \textsc{Metacalibration} adopts galaxy images themselves to relate the measured ellipticity of galaxies to the true shear through the 2$\times$2 response matrix, $\mathbf{R}$. 
The response matrix is calculated by deconvolving the point spread function (PSF) from the image, injecting a small artificial shear and re-convolving the image with the representation of the PSF. The resultant representation of the mean true shear, $\langle \mathbf{\gamma} \rangle$, can be written as, 
\begin{equation}
    \langle \mathbf{\gamma} \rangle = \langle \mathbf{R} \rangle^{-1} \langle \mathbf{e} \rangle.
\end{equation}
In practice, we define the average response as $\Tilde{R}=(\mathbf{R}_{11}+\mathbf{R}_{22})/2$.
We have checked that given the noise level in our data, using this approximation does not affect our measurement significantly.

In addition, there is a second component that contributes to the response matrix, which is due to the selection of the galaxies, $R_{\rm sel}$.
Since the selection response is only meaningful as ensembles of galaxies, we make use of the mean value $\langle R_{\rm sel} \rangle$. 
For details of \textsc{Metacalibration}, we refer the readers to \citet{Sheldon2017}.

In \citet{McClintock2019}, it is shown that the optimal estimator for $\Delta\Sigma (R)$, including the response is,
\begin{equation}
    \widetilde{\Delta\Sigma} (R_{\rm k})=\mathcal{B}(R_{\rm k})\frac{\sum_{\rm i,j}\omega_{\rm i,j} e_{\rm T;i,j}}{\sum_{\rm i,j}\omega_{\rm i,j} \Sigma^{'-1}_{\rm \text{crit};i,j} (\tilde{R_{\rm i}}+\langle R_{\rm sel} \rangle)} \bigg\rvert_{R_{\rm k} < R \leq R_{\rm k+1}},
\end{equation}
for the k-th radial bin, where $\mathcal{B}(R_{\rm k})$ is the correction factor for contamination from the cluster members and foreground galaxies (boost factor), which we describe in the next paragraph.
The summation goes over all the lens (j) - source (i) pairs, and
\begin{equation}
    \Sigma^{'-1}_{\rm \text{crit};i,j} = \Sigma_{\rm \text{crit}}^{-1} (z_{\rm l_{\rm j}}, z_{\rm s_{\rm i}}^{\rm MC}),
\end{equation}
where $z_{\rm s_{\rm i}}^{\rm MC}$ is a random Monte Carlo sample from the full photo-z probability distribution for the i-th source and 
\begin{equation}
    \omega_{\rm i,j} = \Sigma_{\rm \text{crit}}^{-1} (z_{\rm l_{\rm j}}, \langle z_{\rm s_{\rm i}} \rangle) \,\,\,\, \rm{if} \langle z_{\rm s_{\rm i}} \rangle > z_{\rm l_{\rm j}} + 0.1
\end{equation}
for which the photometric redshifts of galaxies are estimated with Directional Neighbourhood Fitting (DNF) algorithm \citep{DeVicente2016}.

Even with a redshift cushion of 0.1 between the lens and the source, because of photometric redshift uncertainties and contamination from the cluster members, some of the source galaxies we use are in front of the lens clusters. 
These galaxies do not retain any gravitational shear due to the lens, therefore dilute the weak lensing signal. 
We correct for this effect following the procedure
in \citet{Sheldon2004},
\begin{equation}
    \mathcal{B}(R_{\rm k}) = \frac{N_{\rm rand}}{ N_{\rm lens}} \frac{\sum_{\rm i,j} \omega_{\rm i,j}}{\sum_{\rm k,l} \omega_{\rm k,l}},
\end{equation}
where i and j represent the lens-source pairs, and k,l the random-source pairs. 

In Figure~\ref{fig:wl}, we show the WL measured surface mass density profiles of the four cluster richness subsets used in this paper. These measurements will be used for direct comparison to the diffuse light radial profiles. The validation cross-component shear measurements around the same clusters, are also shown in Figure~\ref{fig:wl}, which is consistent with a null signal and indicates the shear measurements to be relatively bias-free. Multiplicative bias due to shear calibration or redshift calibration bias may still be present, but will not affect the conclusions of the paper as we do not compare the absolute amplitudes of the lensing and diffuse light luminosity measurements.

\subsection{Conversion into annular surface differential density}
\label{sec:mass_wl}

\begin{figure}
	\includegraphics[width=\columnwidth]{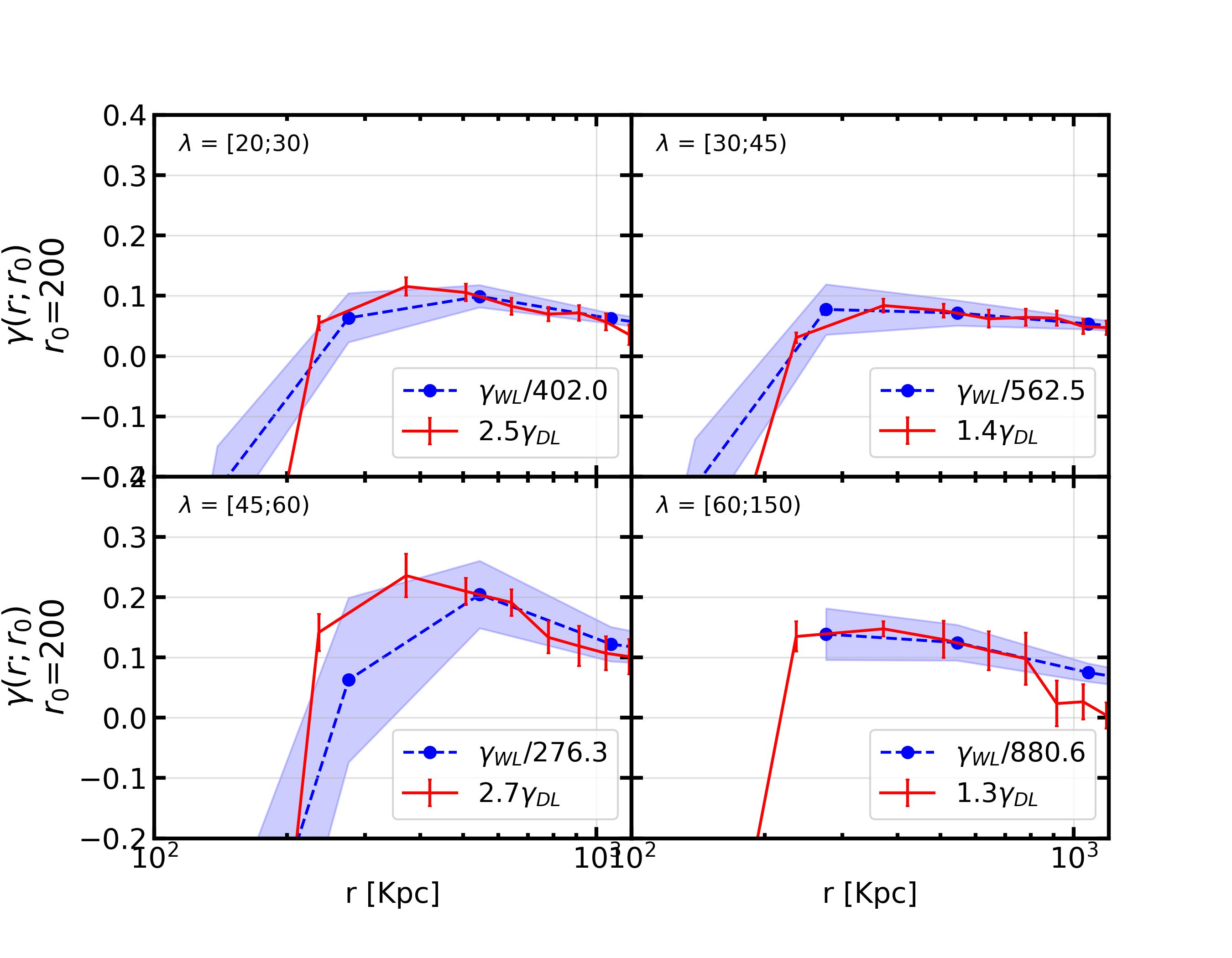}
	\includegraphics[width=\columnwidth]{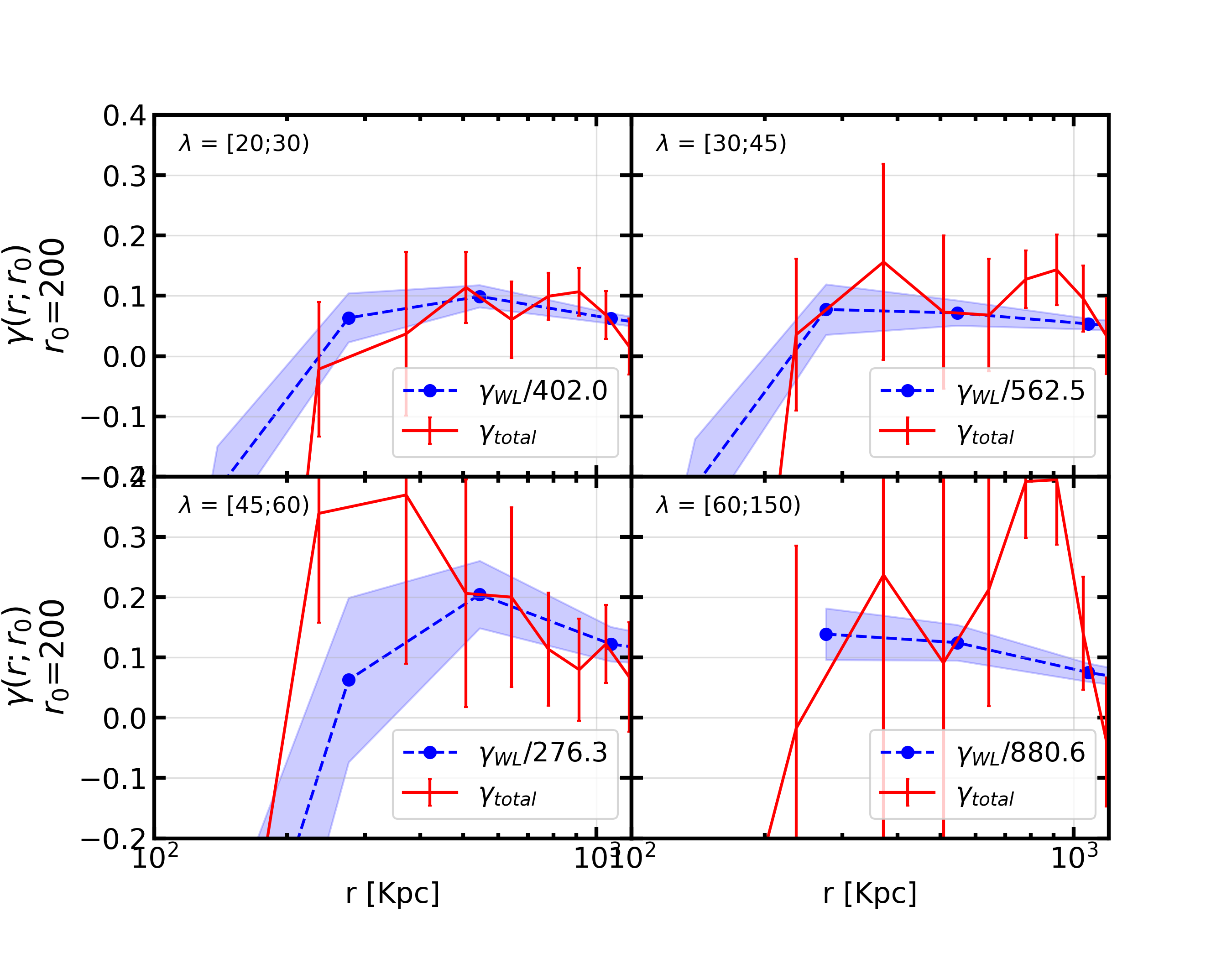}
    \caption{Comparison between the $\Upsilon_\mathrm{DL}$ and $\Upsilon_\mathrm{WL}$ profiles (\textbf{upper panels}) and the $\Upsilon_\mathrm{total}$ and $\Upsilon_\mathrm{WL}$ profiles (\textbf{lower panels}). The red solid lines represent the $\Upsilon_\mathrm{DL}$ and $\Upsilon_\mathrm{total}$ profiles while the blue dashed lines represent the $\Upsilon_\mathrm{WL}$ profiles. The uncertainties of the profiles are derived with the jackknife sampling method. Resemblance between diffuse light ADSB, cluster total light ADSB and cluster total mass ADSD profiles is seen in this plot, and diffuse light seems to trace the cluster total matter distribution beyond 300 kpc closer than cluster total light.}
    \label{fig:icl_wl_comp}
\end{figure}

\begin{figure}
	\includegraphics[width=\columnwidth]{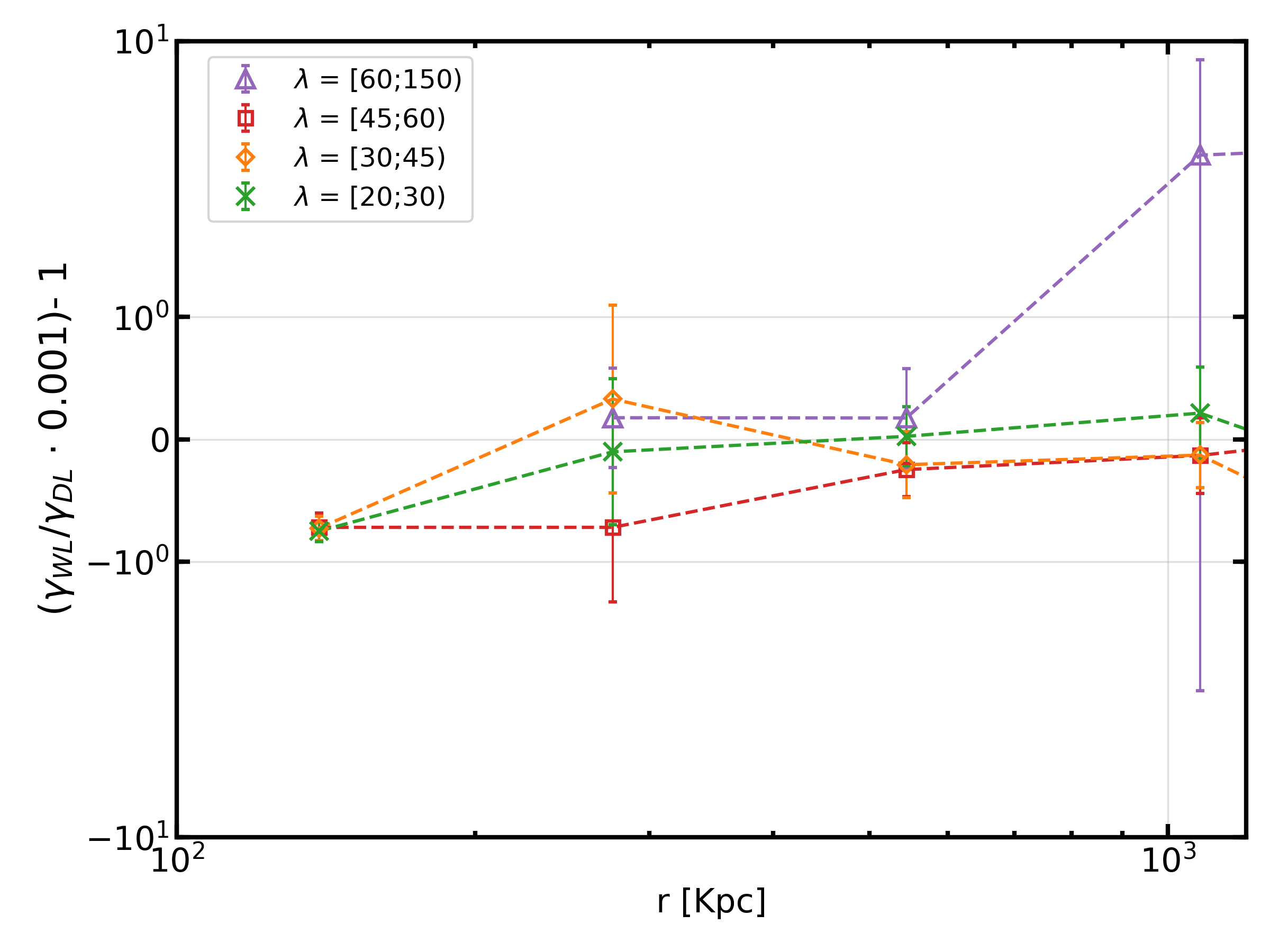}
	\includegraphics[width=\columnwidth]{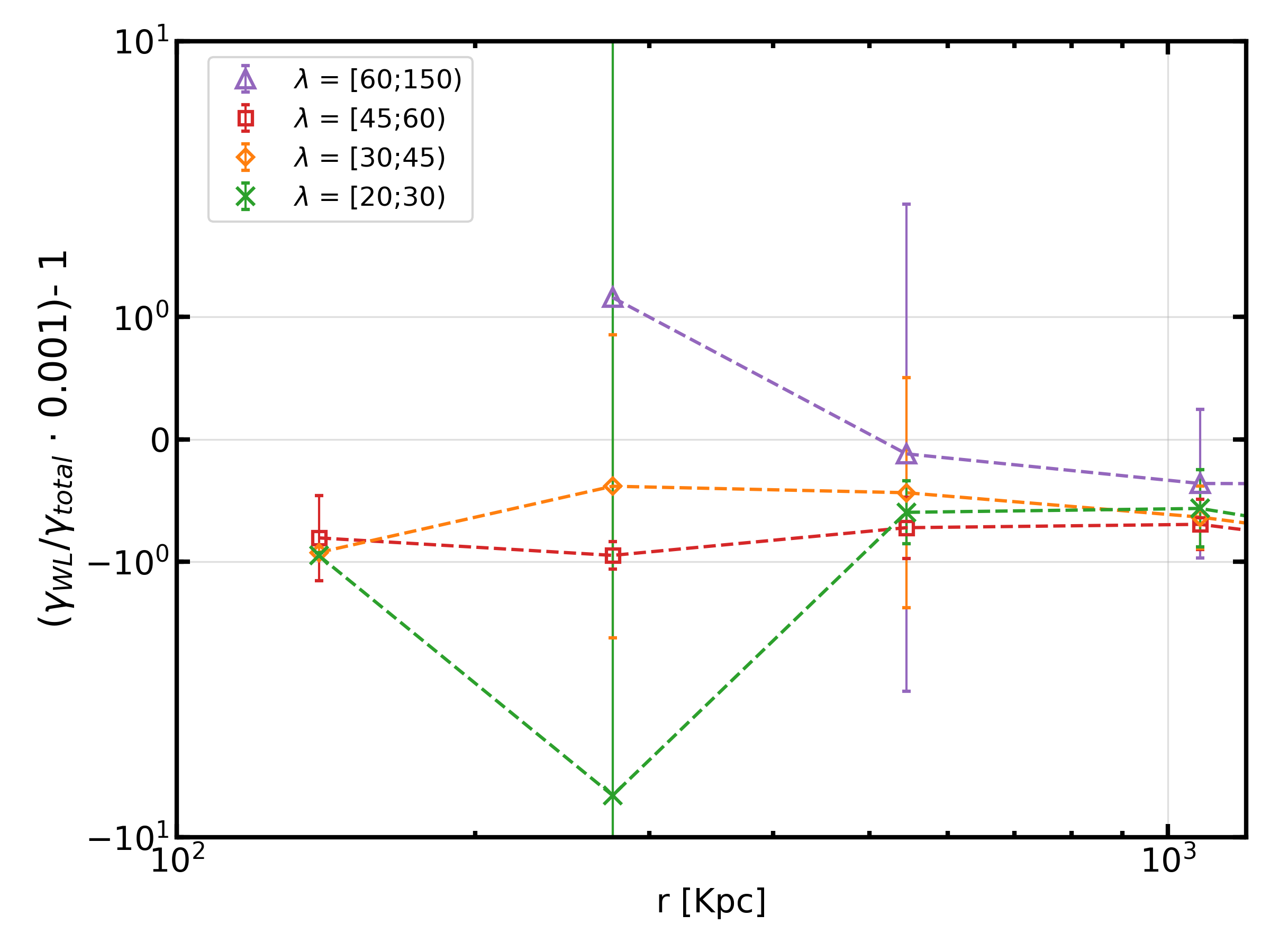}
    \caption{{\bf Upper panel:} Ratios between $\Upsilon_\mathrm{WL}$ and $\Upsilon_\mathrm{DL}$ as a function of cluster radius. {\bf Lower panel:} Ratios between the $\Upsilon_\mathrm{WL}$ and $\Upsilon_\mathrm{total}$ as a function of cluster radius. Note that the y-axes of the two panels use a combination of linear and log scales, linear within -1 to 1, transitioning into log scales outside of 1 or -1 to show large deviations. }
    \label{fig:icl_wl_total_wl_ratio}
\end{figure}

We aim to directly compare the stacked diffuse light radial profiles to the stacked cluster mass profiles measured through weak lensing. However before we start, we need to carefully evaluate what observational quantities to use for such comparison. 

For the diffuse light profiles, we directly measure their surface brightness as a function of radius, which informs us about the diffuse light surface stellar mass density on the plane of the sky. Therefore, it would have been ideal to compare this quantity to the surface mass density of galaxy clusters, $\Sigma(r)$. However, in DES weak lensing measurements (Sec.~\ref{sec:wl-measurements}), the direct observable is the cluster tangential shear profile, which probes the cluster's differential surface mass density, $\Delta\Sigma(r)$,  and is related to the surface mass density $\Sigma(r)$ as, 
\begin{equation}
    \Delta\Sigma(R) = \frac{2}{R^2}\int_0^R R' \Sigma (R') \mathrm{d} R'- \Sigma (R).
	\label{eq:eq_ds}
\end{equation}
Figure \ref{fig:wl} shows the cluster $\Delta\Sigma(r)$ profiles derived from weak lensing in each of the richness sub-samples. 

Although it is possible to derive $\Sigma(r)$ from the weak lensing-measured $\Delta\Sigma(r)$ as is done in \citetalias{Zhang2019}, these derivations rely on model assumptions of the cluster mass distribution. To avoid our diffuse light-weak lensing comparison being affected by model choices, we have decided instead to convert the diffuse light and cluster total light surface brightness into a differential surface brightness as
\begin{equation}
    \Delta L(R) = \frac{2}{R^2}\int_0^R R' L (R') \mathrm{d} R'- L (R).
	\label{eq:eq_dl}
\end{equation}

Note though that cluster differential surface mass density $\Delta\Sigma(R)$ is inevitably affected by the $\Sigma(R)$ values at small radius, where the diffuse light profiles have been shown to have significantly different radial slopes than the cluster total mass distribution in \citetalias{Zhang2019}. To eliminate the small radial contributions, we further convert $\Delta\Sigma(R)$ into the {\bf A}nnular {\bf D}ifferential {\bf S}urface {\bf D}ensity ({\bf ADSD}: \citealt{Mandelbaum2013}), $\Upsilon$, as 
\begin{equation}
    \begin{split}
    \Upsilon_\mathrm{WL}(R;R_{0}) & = \Delta \Sigma (R) - \left( \frac{R_{0}}{R} \right) ^{2} \Delta \Sigma (R_{0}).
    \end{split}
	\label{eq:eq_gamma}
\end{equation}
In the above equation, $R_0$ is a chosen radius within which the cluster's surface mass density will not affect the measurements of $\Upsilon(R;R_{0})$. In this paper we use a $R_0$ value of 200 kpc.
Similarly, we convert the diffuse light and cluster total light differential surface brightness into {\bf A}nnular {\bf D}ifferential {\bf S}urface {\bf B}rightness (ADSB) as 
\begin{equation}
    \begin{split}
    \Upsilon_{\mathrm{DL}/\mathrm{total}}(R;R_{0}) & = \Delta L_{\mathrm{DL}/\mathrm{total}} (R) - \left( \frac{R_{0}}{R} \right) ^{2} \Delta L_{\mathrm{DL}/\mathrm{total}} (R_{0}),
    \end{split}
	\label{eq:eq_gammad}
\end{equation}
 where the sub-script $\mathrm{DL}/\mathrm{total}$ means that the equation applies to both diffuse (DL) and total light of the galaxy cluster.

\subsection{Comparison Result}

In Figure  \ref{fig:icl_wl_comp}, we show the comparisons between the WL-derived cluster mass ADSD and the ADSB of diffuse light, as well as the comparison between the cluster total mass ADSD and the ADSB of cluster total light. Note that the values of the WL-derived cluster ADSD profiles are scaled by the average weak lensing and total-light ratio, ADSD/ADSB, between 550 and 1050 kpc, and the values of the diffuse light ADSB profiles are also scaled by the average diffuse/total-light  ratio between 550 and 1050 kpc, so the cluster ADSD(B) profiles are in similar numerical ranges. 

Overall, we find that the ADSB profiles of diffuse light and the ADSD profiles of cluster total mass have similar radial dependence especially outside 200 kpc, consistent within their 1 $\sigma$ measurement uncertainty range.
However, the ADSB or ADSD profiles, within 200 kpc, start to show some deviations, but the deviation is not significant, and the profiles are still consistent within 1$\sigma$.
Interestingly, the ADSB or ADSD profiles of cluster total light and cluster mass are also consistent within their $1\sigma$ uncertainty ranges, although the ADSB profiles of cluster total light are measured to be much noisier than the ADSB profiles of cluster diffuse light. 
We further derive the ratios between $\Upsilon_\mathrm{WL}$ profiles and $\Upsilon_\mathrm{DL}$ as well as between $\Upsilon_\mathrm{WL}$ and $\Upsilon_\mathrm{total}$, as shown in
Figure  \ref{fig:icl_wl_total_wl_ratio}. Again, we note that the ADSB(D) profiles of diffuse light and cluster total mass have consistent radial dependence outside 200 kpc, but show deviations at a low S/N within 200 kpc. The ADSB(D) profiles of cluster total light and cluster total mass also appear to have consistent radial dependence, but the comparisons are much noisier.

Given these comparisons, we conclude that we see evidence of consistency between diffuse light and cluster total mass radial distributions from weak lensing measurements especially outside 200 kpc of the cluster center. However, given the large uncertainties associated with the ADSD(B) observables, further high S/N measurements of both the cluster weak lensing signals and diffuse light surface brightness will be necessary to distinguish any  subtle differences.  
We will return to this topic of radial resemblance between cluster diffuse light and total mass distribution in Sec.~\ref{sec:sim_prof}.

\section{Diffuse Light Properties in Simulation}
\label{sec:Simulationresult}

In the previous section, we notice similarities between the diffuse light radial profiles and the cluster total mass radial profiles, but can not draw a conclusive statement about their consistency. Diffuse light simulations offer more insight into this aspect. In this section, we turn to the Illustris \textbf{T}he \textbf{N}ext \textbf{G}eneration (IllustrisTNG) hydrodynamic simulation \citep{Nelson2019,Pillepich2018,Springel2018,Nelson2018,Naiman2018,Marinacci2018} to investigate the similarity between the distributions of the diffuse light and the cluster total mass \citep{Pillepich2018}. The IllustrisTNG simulation is a powerful, high-resolution hydrodynamic cosmological simulation, which considers gravitational and hydrodynamic effects as well as sophisticated models for processes of radiation, diffuse gas, and magnetic field. 

We use the IllustrisTNG 300-1 simulation and in particular, the snapshot at redshift 0.27, which matches the median redshift of our redMaPPer cluster samples. We select halos with masses $M_{200\text{m}} \gtrapprox 7.5 \times10^{13} \mathrm{M_\odot/h}$, which roughly matches the mass range of the redMaPPer clusters analyzed in this paper, and eliminate halos that are within 20 Mpc/$h$ of the snapshot boundaries to avoid boundary effects. These selection criteria yield 110 halos suitable for our analysis. We then derive the densities of the simulation stellar particles, dark matter particles, and gaseous particles as a function of 3D halo radius, and also 2D projected radius on the simulation $x/y$ plane. Note though the stellar distribution in the IllustrisTNG 300-1 simulation has not fully converged, which is limited by the simulation resolution as studied in \cite{Pillepich2018}. The simulation convergence issue affects the total stellar mass in halos and the shape of the halo stellar mass radial profile within the central 2-4 kpc .  \cite{Pillepich2018} has rescaled the TNG 300-1 stellar mass results using the smaller volume but higher resolution TNG 100-1 simulation, but we do not rescale our results based on TNG 300-1 given the small number of halos in our interested mass range in the TNG 100-1 simulation.

\subsection{Does diffuse light have the same radial dependence with cluster total mass?}
\label{sec:sim_prof}

\begin{figure}
	\includegraphics[width=\columnwidth]{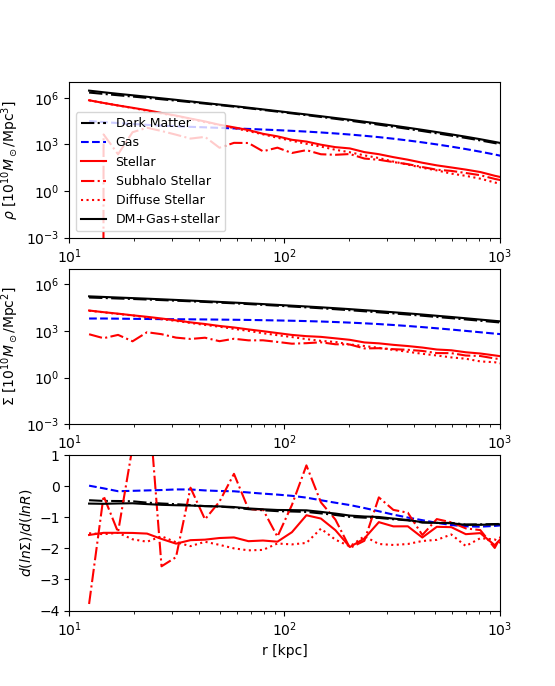}
    \caption{The radial density profiles in 3D ({\bf upper panel}) and 2D projected distances ({\bf middle panel}) of various cluster mass components in the IllustrisTNG 300-1 simulation -- dark matter, gas, diffuse light and subhalo stellar mass -- as well as the total halo mass densities. {\bf Lower panel:} The radial derivative of the radial densities in 2D projected space. Throughout the plots, we notice that the halo diffuse light appears to be the most concentrated, while the halo gaseous content appears to have the lowest radial concentration, which is consistent with diffuse light being produced from galaxy stripping/disruption towards the halo center, while gaseous particles experience frequent interactions that flatten out their radial distribution. The most faithful radial tracer of the halo dark matter distribution appears to be the subhalo stellar mass.}
    \label{fig:tng-profiles}
\end{figure}

\begin{figure}
	\includegraphics[width=\columnwidth]{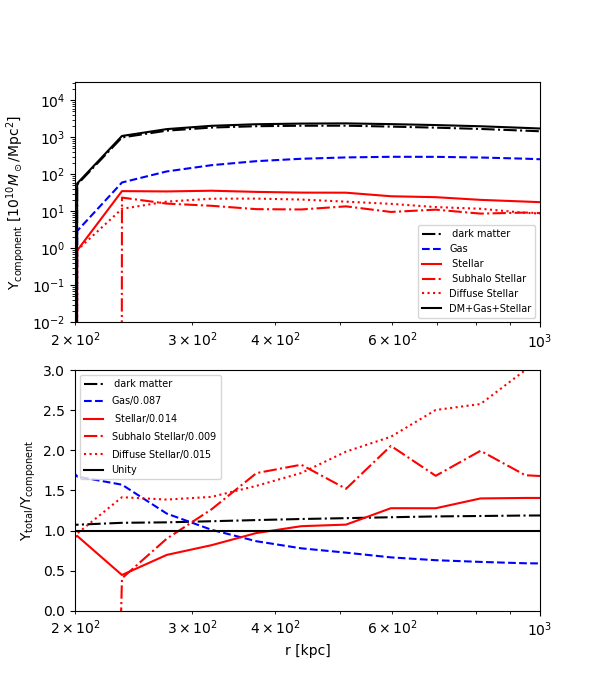}
    \caption{{\bf Upper Panel:} the $\Upsilon$ density profiles of various cluster mass components -- dark matter, gas, diffuse light and subhalo stellar mass -- as well as the total halo mass densities, derived from the 2D projected radial density profiles in Figure~\ref{fig:tng-profiles}. {\bf Lower Panel:} the relative ratios between the $\Upsilon$ profiles of halo total mass and the various halo mass components -- dark matter, gas, diffuse light and subhalo stellar mass. The $\Upsilon$ profiles of the subhalo stellar mass appears to have the same radial trend with the total halo mass profile, while the $\Upsilon$ profile of the halo diffuse light appears to drop more rapidly with increasing radius than the total halo mass $\Upsilon$ profile. The least radially concentrated halo mass component, i.e., the halo gaseous mass, has a $\Upsilon$ profile that drops least rapidly with increasing radius.}
    \label{fig:tng-gamma}
\end{figure}

To derive the radial density profiles of the diffuse light in the simulation, we first compute the radial density profiles of the stellar particles contained in subhalos.  The stellar mass of subhalos within twice the stellar mass half-radius of the subhalos are used to derive these profiles, although we limit the calculation to the subhalos of stellar mass above $10^9 M_\odot$ (contained within the radius of $V_{max}$). The subhalo radius and the subhalo mass thresholds are selected to roughly match the galaxy masking radius and depth limit of our measurements. This subhalo stellar profile is then subtracted from the radial density profiles of all the stellar particles around the halos, and the subtracted result is considered the diffuse stellar radial distribution. These subtractions are done in both 3D and 2D to derive the 3D and projected 2D radial distributions of the diffuse light.

The upper and middle panels of Figure~\ref{fig:tng-profiles} show those radial dark matter, gaseous and stellar profiles, averaged over all the selected halos to reduce noise. In either the 3D radial profiles or the projected 2D radial profiles, the total halo stellar content appears to have the most concentrated radial distribution, while the halo gaseous component appears to have the least concentrated radial distribution due to the high interaction rate between the gaseous particles. Neither the stellar particles nor the gaseous particles appear to faithfully follow the radial dependence of dark matter (or halo total mass). However, after separating the total halo stellar content into the diffuse and the subhalo components, we notice that the subhalo stellar component is following the dark matter (or the halo total mass) radial distribution remarkably well, while the diffuse stellar component deviates further from the halo dark matter radial distribution, and becomes the most radially concentrated halo component. 

The lower panel of Figure~\ref{fig:tng-profiles} shows the 2D radial density derivatives of the various halo components. As noted above, the most faithful radial tracer of the halo dark matter (or the halo total mass) distribution appears to be the subhalo stellar mass. The halo gaseous component has the mildest radial slope among all of the analyzed components, while the halo diffuse stellar component has the steepest radial slope and thus is the most radially concentrated. Since diffuse light is expected to originate from galaxy stripping/disruption, which can only happen at small halo radii after the sub-halos's outer dark matter component is completely destroyed, these simulation findings are not particularly surprising, if not limited by the relaxation timescale of the diffuse light after their origination.

We further convert the simulation projected 2D radial densities into a $\Upsilon$ radial profile, so as to be more directly comparable to the measured cluster matter/diffuse light density profiles in Section~\ref{sec:mass_wl}. The conversion made it less obvious to directly spot the radial concentration of the various halo components as shown in the upper panel of Figure~\ref{fig:tng-gamma}, thus we plot the ratios between the various component's $\Upsilon$ profile to the $\Upsilon$ profile of the total halo mass. The $\Upsilon$ profile of halo diffuse light drops most quickly with halo radius, while the least concentrated halo gaseous component has a $\Upsilon$ profile that drops the least quickly with halo radius. The $\Upsilon$ profile of the subhalo stellar mass appears to have consistent radial trend with the total halo mass.  
We conclude that the simulation results do not support that diffuse light is a faithful radial tracer of the cluster total mass, although cluster satellite galaxy stellar content is.

We note that in this $\Upsilon$ profile comparison, the radial dependencies of the various halo components appear to be only distinguishable when the measurements are made with high S/N. Given that the $\Upsilon$ ratios between diffuse/total stellar and cluster mass change, at most, by a factor of $\sim 3$ from 200 to 1000 kpc, our observational-based measurements in Sec.~\ref{sec:mass_wl} likely does not have enough signal-to-noise to distinguish dissimilarity of radial trends between diffuse light, cluster total light, and total mass -- in the future, higher S/N measurements  of cluster weak lensing signals as well as light distributions are necessary to confirm our findings in observations.

\subsection{Is diffuse light a good indicator of cluster mass?}
\label{sec:sim_mass_indicator}

\begin{figure*}
	\includegraphics[width=2.2\columnwidth]{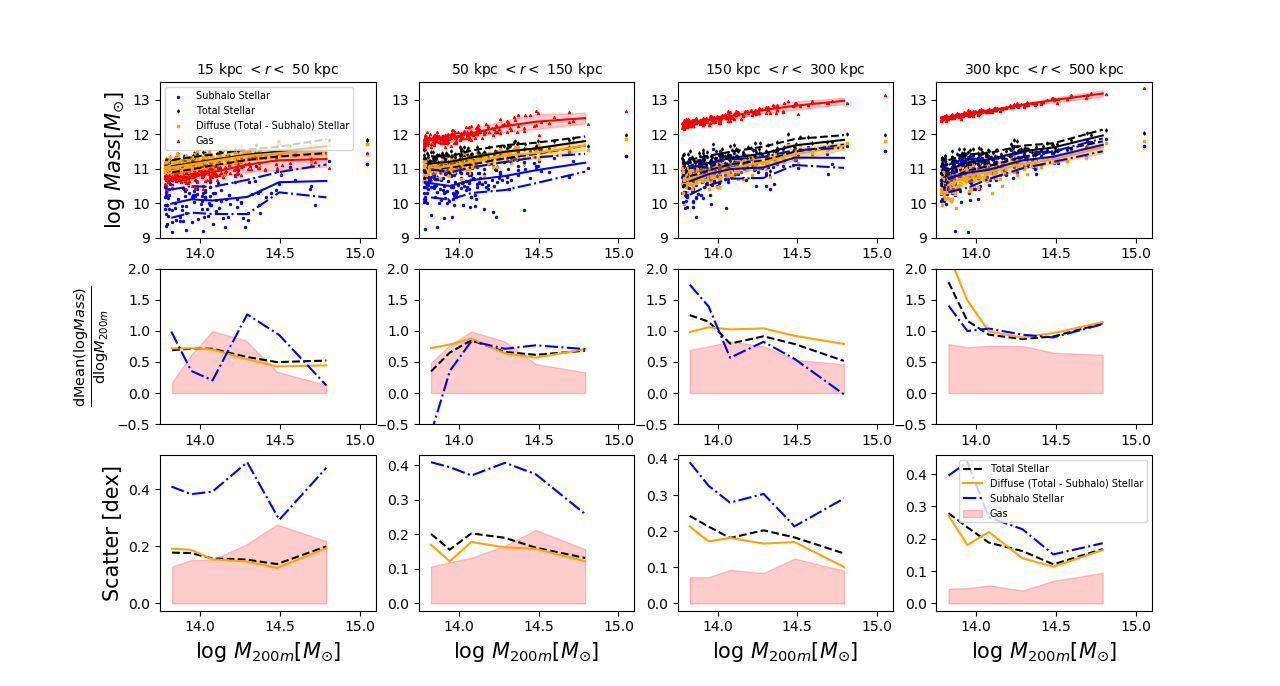}
    \caption{{\bf Upper panel:} Halo mass components integrated in four radial ranges {\it VS} halo total mass $M_{200\text{m}}$ in the IllustrisTNG 300-1 simulation. {\bf Middle panel:} Slopes of the mean relations between the various halo mass components and cluster total mass. {\bf Lower panel:} Scatter of the various halo mass components in fixed halo mass ranges, which shows that the halo gaseous component has the least scattered halo mass indicator, while the halo diffuse light also appears to be a reasonable halo mass indicator with relatively low scatter.}
    \label{fig:tng-tracer}
\end{figure*}

In Sec. ~\ref{sec:sim_prof}, we find that simulation diffuse light is not a faithful radial tracer of the cluster matter distribution, but our analysis as well as previous studies have clearly noted a strong correlation between diffuse light and the  cluster's total mass \citep[e.g.][]{Pillepich2014, Pillepich2018, Montes2018}, such as the similar shape in the radial density contour lines between diffuse light and cluster mass \citep[e.g.][]{ Montes2019, Asensio2020} and the self-similarity of the diffuse light radial profiles (\citetalias{Zhang2019}). It is possible that although diffuse light does not faithfully trace the cluster matter distribution, it simply follows a different radial distribution that still has a strong dependence on cluster total mass. Thus, even if the diffuse light profiles can not be used to directly map out the dark matter distribution inside clusters, its total luminosity can still serve as a strong cluster mass indicator. 

In this subsection, we examine the correlations between halo mass and the various halo baryonic mass components, including the diffuse light, the sub-halo stellar mass and the gaseous mass. For each halo in the simulation that are at least 20 Mpc/$h$ away from the simulation boundaries (to avoid the results being affected by the boundary of the simulation), we derive their diffuse stellar masses, subhalo stellar masses, total stellar masses (diffuse + subhalo) and gaseous masses, integrated over 3D radial ranges. The relations between the masses of those components and the cluster's total mass is shown in Figure~\ref{fig:tng-tracer}. 

In the radial ranges above 50 kpc, all the cluster baryonic mass components show clear correlations with cluster mass. From 15 kpc to 50 kpc, the cluster subhalo stellar mass do not show significant correlation with cluster mass; the diffuse and the gaseous mass still show correlations, but the mass dependence is milder than the other radial ranges as measured by the slope of the component-mass/halo-mass relations. 

A particularly interesting quantity is the scatters of these component-masses at fixed halo mass. In the lower panel of Figure~\ref{fig:tng-tracer}, we show the mean scatter of the component-masses around their mean values in a fixed halo mass range. As well known \citep[e.g.,][]{Voit2005, Kravtsov2006} in previous studies, the halo gaseous mass is an excellent low-scatter indicator of halo mass, showing the lowest scatter in our examination --around 0.1 dex throughout the 15 kpc to the 500 kpc radial range. The diffuse stellar mass, appears to be the next best low-scatter mass indicator with a scatter around 0.2 to 0.25 dex in the radial range of 15 kpc to 300 kpc. However, the scatter of the diffuse stellar mass does increase with radius caused by the rapid decrease of diffuse stellar density with radius. The halo total stellar mass has consistent scatter with the halo diffuse mass within 300 kpc, but this is likely due to the domination of halo diffuse light in the total halo stellar content within this radius range. The subhalo stellar mass has the highest scatter among all of the probed components, around 0.2 to 0.5 dex depending on the radial or halo-mass range. Outside of 300 kpc, subhalo stellar mass starts to have similar scatter with the diffuse mass, and meanwhile becomes a bigger contributor to the halo total stellar mass over the diffuse stellar mass. 

Comparing those stellar mass components, we highly recommend using halo diffuse stellar mass, or halo total stellar mass within 500 kpc as a robust, low-scatter halo mass indicator. The halo total stellar mass estimation must include halo diffuse light to minimize the scatter within 300 kpc, which has not been studied in previous analyses \citep{2020MNRAS.493.4591P, 2020arXiv200102283A}. This simulation analysis conclusion is also in agreement with our observational result in Sec.~\ref{sec:fluxprof}, in which we find strong correlation between diffuse light luminosity and cluster total mass, which is even more evident than the correlation between cluster total light and mass (Sec.~\ref{sec:tot_l_prof}).

Note though, these simulation conclusions are derived with halo components mass enclosed within 3D radii, while observations are almost always measured in 2D projected radii, and thus affected by foreground and background structures. We find that using halo stellar mass enclosed within 2D projected radii increases its scatter, but it may be possible to reduce such a scatter in real observations with imaging colour information (we note that the satellite galaxies in massive halos in the simulation display broader color distributions than observations). Given the vital importance of developing low-scatter halo-mass indicators in cluster cosmological studies, it would be interesting to carry out observational studies of cluster total stellar mass or cluster diffuse stellar mass, especially using multi-wavelength data that can observationally evaluate the scatter of  cluster mass indicators \citep[e.g.,][]{Farahi2019, 2020MNRAS.493.4591P}. 

\subsection{Additional diffuse light properties}

\begin{figure}
	\includegraphics[width=1\columnwidth]{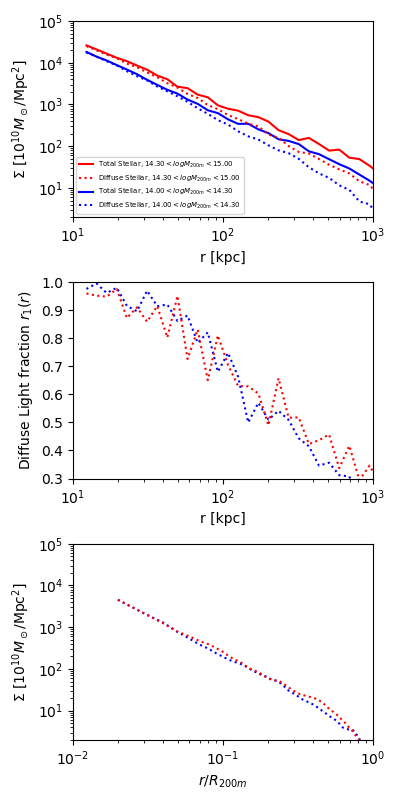}
    \caption{Diffuse stellar mass properties in the simulation: the diffuse stellar mass and cluster total stellar mass radial profiles ({\bf upper panel}), the diffuse stellar fraction ({\bf middle panel}) and the cluster $R_{200\text{m}}$-scaled diffuse stellar mass and cluster total stellar mass radial profiles ({\bf bottom panel}). These diffuse stellar properties as well as the cluster total stellar properties are in qualitative agreement with the observational measurements in this paper.}
    \label{fig:tng-df}
\end{figure}

As a qualitative comparison to the observational results presented in Sec.~\ref{sec:massdep}, we derive additional diffuse stellar mass properties in the IllustrisTNG 300-1 simulation and show those in Figure ~\ref{fig:tng-df}. We demonstrate the 2D-projected radial profiles of the diffuse stellar mass and the halo total stellar mass, in two halo mass ranges (limited by the small size of the IllustrisTNG cluster sample), and then derive the ratios between the two as the diffuse stellar fraction in the simulation. 

These two results are qualitatively comparable to the observational results shown in Figure~\ref{fig:iclprofiles_lup},~\ref{fig:mag_unm_lup_comparison} and~\ref{fig:icl_frac}. We find that diffuse stellar/light is more abundant in more massive clusters, and the diffuse stellar/light fractions do not appear to change with cluster mass. However, the diffuse stellar fractions appear to be significantly higher in the simulation than in observations, as high as $\sim$40\% at 1 Mpc of the cluster center, while the observational measurements are around $\sim$ 30\%. It is possible that diffuse light has been over-produced in the simulation. 

We also averaged the diffuse stellar mass profiles after scaling by cluster radius ($R_{\mathrm{200m}}$). In good agreement with our observational finding (Sec.~\ref{sec:self_similarity}), the diffuse stellar mass profiles also display self-similarity, that their radial profiles appear to be uniform after scaling by cluster radius $R_{\mathrm{200m}}$.

\section{Conclusions}
\label{sec:conclusions}

In this paper, we present for the first time a direct comparison of the radial dependence of the diffuse light surface brightness and the weak-lensing measured cluster matter distribution, for a statistically large cluster sample with high S/N diffuse light measurements to a cluster radial range of 1 Mpc. We also present both observational and simulation evidence for a strong correlation of diffuse light luminosity with cluster mass. The findings can be summarized as the following.
\begin{itemize}
  \item \textit{Strong correlation between diffuse light brightness and cluster mass at large radius:} We observe that more massive clusters have more diffuse light in the regions outside  20 kpc of the cluster center, and the mass dependence becomes steeper with increasing cluster radius.   
  The total stellar luminosities contained within 15 kpc of the cluster centers are almost indistinguishable between clusters of different richnesses/masses, but the total stellar luminosities contained around $\sim$ 300 kpc of the cluster radius show significant correlation with cluster total mass.   
    \item \textit{Self-similarity of the diffuse light radial profiles:} the diffuse light surface brightness radial profiles appear to have a universal distribution at intermediate and large radius after scaling by cluster $R_\mathrm{200m}$.

  \item \textit{Mass (in)dependence of the diffuse light fraction:} we derive the diffuse light fraction in total cluster stellar luminosity as a function of cluster radius and mass. The cumulative diffuse light fraction drops with enlarging cluster radius, reaching $\sim$ 24\% at $\sim$ 700 kpc. Interestingly, we do not find diffuse light fraction to be dependent on cluster mass within 1 Mpc of the cluster radius, possibly because the cluster growth is well correlated with diffuse light accretion within this radial range.
 
  \item \textit{Comparison to weak lensing matter distribution:} we directly compare the radial density distribution of diffuse light to that of the cluster total matter (including dark matter) measured through weak lensing. We find that the diffuse light radial distributions indeed show some level of resemblance with the cluster matter distributions. In addition, the radial distribution of cluster total stellar mass also appears to have a similar, but noisier similarity with cluster matter. 

  \item \textit{Diffuse light properties in the IllustrisTNG simulation.}  
In the IllustrisTNG simulation, the diffuse light radial distribution is more concentrated towards the center than the cluster mass (including dark matter mass) distribution, while the radial profile of the cluster subhalo stellar mass appears to well match that of the cluster mass. We do find that the total stellar mass of diffuse light at large radii scales remarkably well with the cluster mass with a low scatter, comparable to the scaling relation of cluster gaseous mass within 150 kpc, and outperforms the cluster subhalo stellar mass throughout the 0 to 500 kpc radial range. This result is consistent with our observation that diffuse light has an excellent scaling relation with cluster mass. 

\end{itemize}

Given our results, is diffuse intracluster light a good tracer of the galaxy cluster matter distribution (including dark matter)? Our answer is maybe. 

Observationally, we find that the diffuse light radial profile shows some resemblance with that of cluster matter measured through weak lensing, but simulation analysis suggests that they are not tracing each other faithfully which needs to be  confirmed with more studies. However, the diffuse light luminosity at large radius scales extraordinarily well with cluster total mass with a power-law like relation in both observation and simulation. We hence recommend developing the diffuse light observable as a potential low scatter mass indicator for  cluster astrophysics and cosmology studies. Such mass proxies can be particularly useful for low-mass clusters where multi-wavelength data is scarce and accurate cluster mass estimation is challenging \citep[e.g., see discussion  in][]{2020arXiv200211124D}, but existing wide-field optical survey programs like DES offer deep enough data to acquire accurate measurements of diffuse light.

Moving forward, these interesting findings can enjoy a better understanding with higher S/N measurements. The next generation of wide-field survey programs such as the Legacy Survey of Space and Time\footnote{\url{https://www.lsst.org}} (LSST) based at the Vera Rubin Observatory and the Euclid Wide Survey\footnote{\url{https://www.euclid-ec.org}} provide great opportunities to further investigate the properties of cluster diffuse light. Moreover, we have not explored the effect of cluster relaxation process on diffuse light production, or studied the correlation between cluster morphological parameters (smoothness, cuspiness, asymmetry, and concentration) and diffuse light. Meanwhile, simulation studies still need to explain the origin of diffuse light and present evidence that matches diffuse light properties in observations. We advocate that continuing to study diffuse light with both observations and simulations will have much to contribute to understanding galaxy and galaxy cluster evolution.

\section*{Acknowledgements}
\label{sec:ack}

We thank Amanda Pagul, Joe Mohr and Chihway Chang for helpful discussions.
Funding for the DES Projects has been provided by the U.S. Department of Energy, the U.S. National Science Foundation, the Ministry of Science and Education of Spain, 
the Science and Technology Facilities Council of the United Kingdom, the Higher Education Funding Council for England, the National Center for Supercomputing 
Applications at the University of Illinois at Urbana-Champaign, the Kavli Institute of Cosmological Physics at the University of Chicago, 
the Center for Cosmology and Astro-Particle Physics at the Ohio State University,
the Mitchell Institute for Fundamental Physics and Astronomy at Texas A\&M University, Financiadora de Estudos e Projetos, 
Funda{\c c}{\~a}o Carlos Chagas Filho de Amparo {\`a} Pesquisa do Estado do Rio de Janeiro, Conselho Nacional de Desenvolvimento Cient{\'i}fico e Tecnol{\'o}gico and 
the Minist{\'e}rio da Ci{\^e}ncia, Tecnologia e Inova{\c c}{\~a}o, the Deutsche Forschungsgemeinschaft and the Collaborating Institutions in the Dark Energy Survey. 

The Collaborating Institutions are Argonne National Laboratory, the University of California at Santa Cruz, the University of Cambridge, Centro de Investigaciones Energ{\'e}ticas, 
Medioambientales y Tecnol{\'o}gicas-Madrid, the University of Chicago, University College London, the DES-Brazil Consortium, the University of Edinburgh, 
the Eidgen{\"o}ssische Technische Hochschule (ETH) Z{\"u}rich, 
Fermi National Accelerator Laboratory, the University of Illinois at Urbana-Champaign, the Institut de Ci{\`e}ncies de l'Espai (IEEC/CSIC), 
the Institut de F{\'i}sica d'Altes Energies, Lawrence Berkeley National Laboratory, the Ludwig-Maximilians Universit{\"a}t M{\"u}nchen and the associated Excellence Cluster Universe, 
the University of Michigan, the National Optical Astronomy Observatory, the University of Nottingham, The Ohio State University, the University of Pennsylvania, the University of Portsmouth, 
SLAC National Accelerator Laboratory, Stanford University, the University of Sussex, Texas A\&M University, and the OzDES Membership Consortium.

Based in part on observations at Cerro Tololo Inter-American Observatory, National Optical Astronomy Observatory, which is operated by the Association of 
Universities for Research in Astronomy (AURA) under a cooperative agreement with the National Science Foundation.

The DES data management system is supported by the National Science Foundation under Grant Numbers AST-1138766 and AST-1536171.
The DES participants from Spanish institutions are partially supported by MINECO under grants AYA2015-71825, ESP2015-66861, FPA2015-68048, SEV-2016-0588, SEV-2016-0597, and MDM-2015-0509, 
some of which include ERDF funds from the European Union. IFAE is partially funded by the CERCA program of the Generalitat de Catalunya.
Research leading to these results has received funding from the European Research
Council under the European Union's Seventh Framework Program (FP7/2007-2013) including ERC grant agreements 240672, 291329, and 306478.
We  acknowledge support from the Brazilian Instituto Nacional de Ci\^encia
e Tecnologia (INCT) e-Universe (CNPq grant 465376/2014-2).

This manuscript has been authored by Fermi Research Alliance, LLC under Contract No. DE-AC02-07CH11359 with the U.S. Department of Energy, Office of Science, Office of High Energy Physics.


\section*{Affiliations}
{\small
$^{1}$ Observat\'orio Nacional, Rua Gal. Jos\'e Cristino 77, Rio de Janeiro, RJ - 20921-400, Brazil\\
$^{2}$ Laborat\'orio Interinstitucional de e-Astronomia - LIneA, Rua Gal. Jos\'e Cristino 77, Rio de Janeiro, RJ - 20921-400, Brazil\\
$^{3}$ Fermi National Accelerator Laboratory, P. O. Box 500, Batavia, IL 60510, USA\\
$^{4}$ Department of Physics and Astronomy, University of Pennsylvania, Philadelphia, PA 19104, USA\\
$^{5}$ Department of Astronomy, Shanghai Jiao Tong University, Shanghai 200240, China\\
$^{6}$ Department of Astronomy, University of Michigan, Ann Arbor, MI 48109, USA\\
$^{7}$  Astrophysics \& Cosmology Research Unit, School of Mathematics, Statistics \& Computer Science, University of KwaZulu-Natal, Westville Campus, Private Bag X54001, Durban 4000, South Africa\\
$^{8}$ Center for Cosmology and Astro-Particle Physics, The Ohio State University, Columbus, OH 43210, USA\\
$^{9}$ Institut de F\'{\i}sica d'Altes Energies (IFAE), The Barcelona Institute of Science and Technology, Campus UAB, 08193 Bellaterra (Barcelona) Spain\\
$^{10}$ Department of Physics, Stanford University, 382 Via Pueblo Mall, Stanford, CA 94305, USA\\
$^{11}$ Kavli Institute for Particle Astrophysics \& Cosmology, P. O. Box 2450, Stanford University, Stanford, CA 94305, USA\\
$^{12}$ SLAC National Accelerator Laboratory, Menlo Park, CA 94025, USA\\
$^{13}$ Max Planck Institute for Extraterrestrial Physics, Giessenbachstrasse, 85748 Garching, Germany\\
$^{14}$ Universit\"ats-Sternwarte, Fakult\"at f\"ur Physik, Ludwig-Maximilians Universit\"at M\"unchen, Scheinerstr. 1, 81679 M\"unchen, Germany\\
$^{15}$ McWilliams Center for Cosmology, Department of Physics, Carnegie Mellon University, Pittsburgh, Pennsylvania 15312, USA\\
$^{16}$ Centro de Investigaciones Energ\'eticas, Medioambientales y Tecnol\'ogicas (CIEMAT), Madrid, Spain\\
$^{17}$ Institute for Astronomy, University of Edinburgh, Edinburgh EH9 3HJ, UK\\
$^{18}$ Cerro Tololo Inter-American Observatory, National Optical Astronomy Observatory, Casilla 603, La Serena, Chile\\
$^{19}$ Departamento de F\'isica Matem\'atica, Instituto de F\'isica, Universidade de S\~ao Paulo, CP 66318, S\~ao Paulo, SP, 05314-970, Brazil\\
$^{20}$ Instituto de Fisica Teorica UAM/CSIC, Universidad Autonoma de Madrid, 28049 Madrid, Spain\\
$^{21}$ CNRS, UMR 7095, Institut d'Astrophysique de Paris, F-75014, Paris, France\\
$^{22}$ Sorbonne Universit\'es, UPMC Univ Paris 06, UMR 7095, Institut d'Astrophysique de Paris, F-75014, Paris, France\\
$^{23}$ Department of Physics \& Astronomy, University College London, Gower Street, London, WC1E 6BT, UK\\
$^{24}$ Department of Astronomy, University of Illinois at Urbana-Champaign, 1002 W. Green Street, Urbana, IL 61801, USA\\
$^{25}$ National Center for Supercomputing Applications, 1205 West Clark St., Urbana, IL 61801, USA\\
$^{26}$ Department of Astronomy and Astrophysics, University of Chicago, Chicago, IL 60637, USA\\
$^{27}$ Kavli Institute for Cosmological Physics, University of Chicago, Chicago, IL 60637, USA\\
$^{28}$ INAF-Osservatorio Astronomico di Trieste, via G. B. Tiepolo 11, I-34143 Trieste, Italy\\
$^{29}$ Institute for Fundamental Physics of the Universe, Via Beirut 2, 34014 Trieste, Italy\\
$^{30}$ Santa Cruz Institute for Particle Physics, Santa Cruz, CA 95064, USA\\
$^{31}$ Department of Physics, University of Michigan, Ann Arbor, MI 48109, USA\\
$^{32}$ Institut d'Estudis Espacials de Catalunya (IEEC), 08034 Barcelona, Spain\\
$^{33}$ Institute of Space Sciences (ICE, CSIC),  Campus UAB, Carrer de Can Magrans, s/n,  08193 Barcelona, Spain\\
$^{34}$ School of Mathematics and Physics, University of Queensland,  Brisbane, QLD 4072, Australia\\
$^{35}$ Department of Physics, The Ohio State University, Columbus, OH 43210, USA\\
$^{36}$ Center for Astrophysics $\vert$ Harvard \& Smithsonian, 60 Garden Street, Cambridge, MA 02138, USA\\
$^{37}$ Australian Astronomical Optics, Macquarie University, North Ryde, NSW 2113, Australia\\
$^{38}$ Lowell Observatory, 1400 Mars Hill Rd, Flagstaff, AZ 86001, USA\\
$^{39}$ George P. and Cynthia Woods Mitchell Institute for Fundamental Physics and Astronomy, and Department of Physics and Astronomy, Texas A\&M University, College Station, TX 77843,  USA\\
$^{40}$ Instituci\'o Catalana de Recerca i Estudis Avan\c{c}ats, E-08010 Barcelona, Spain\\
$^{41}$ Institute of Astronomy, University of Cambridge, Madingley Road, Cambridge CB3 0HA, UK\\
$^{42}$ Department of Astrophysical Sciences, Princeton University, Peyton Hall, Princeton, NJ 08544, USA\\
$^{43}$ Instituto de F\'\i sica, UFRGS, Caixa Postal 15051, Porto Alegre, RS - 91501-970, Brazil\\
$^{44}$ School of Physics and Astronomy, University of Southampton,  Southampton, SO17 1BJ, UK\\
$^{45}$ Computer Science and Mathematics Division, Oak Ridge National Laboratory, Oak Ridge, TN 37831\\
}

\section*{Data Availability}

The data underlying this article will be shared on reasonable request to the corresponding author.

\bibliographystyle{abc}
\bibliography{references}


\bsp	
\label{lastpage}

\end{document}